\documentclass[amsmath,amssymb,aps,eqsecnum]{revtex4}
\usepackage[dvips]{graphicx}
 \usepackage{bm,bbm}
\usepackage{epsfig}

\begin{document}
\title{On discrete structures in finite Hilbert spaces}
\author{Ingemar Bengtsson$^1$ and Karol {\.Z}yczkowski$^{2,3}$}

\affiliation{$^1$Fysikum, Stockholm University, Sweden}

\affiliation {$^2$Jagiellonian University, Cracow, Poland} 
\affiliation{$^3$Center for Theoretical Physics, 
  Polish Academy of Sciences Warsaw, Poland}

 \date{January 27, 2017}

\begin{abstract}
  We present a brief review of discrete structures in a finite Hilbert space, 
  relevant for the theory of quantum information.
  Unitary operator bases, mutually unbiased bases,
  Clifford group and stabilizer states, discrete Wigner function, 
  symmetric informationally complete measurements,
  projective and unitary t--designs are discussed. Some recent results in the 
  field are covered and several important open questions are
  formulated. We advocate a geometric approach to the subject and
  emphasize numerous links to various mathematical problems 

\end{abstract}

%
\maketitle

\medskip
\begin{center}
{\small e-mail:   ingemar@physto.se \quad karol@cft.edu.pl}
\end{center}


\section{Introduction}

These notes are based on a new chapter written to the second edition
of our book  {\sl Geometry of Quantum States.
An introduction to Quantum Entanglement} \cite{BZ06}.
The book is written 
at the graduate level for a reader familiar with the principles of quantum mechanics.
It is targeted first of all for readers who 
do not read the mathematical literature everyday, but 
we hope that students of mathematics and of the information sciences will find it useful as well, since they also may wish to learn about 
quantum entanglement.

Individual chapters of the book are to a large extent
independent of each other. For instance, we hope
that the new chapter presented here
might become a source of information 
on recent developments on discrete structures 
in finite Hilbert space also for experts working in the field.
Therefore we have compiled these notes, which aim to present
an introduction to the subject as well as an up to date
review on basic features of objects belonging to the Hilbert space
and important for the field of quantum information processing.

Quantum state spaces are continuous, but they have some intriguing 
realizations of discrete structures hidden inside. We will discuss some of 
them, starting from unitary operator bases, a notion of strategic importance 
in the theory of entanglement, signal processing, quantum computation, and 
more. The structures we are aiming at are known under strange acronyms such 
as `MUB' and `SIC'. They will be spelled out in due course, but in most of 
the chapter we let the Heisenberg groups occupy the centre stage. It seems 
that the Heisenberg groups understand what is going on. 

All references to equations or the numbers of section refers
to the draft of the second edition of the book.
To give a reader a better orientation on the topics
covered we provide its contents in Appendix A.
The second edition of the book includes also 
a new chapter 17 on multipartite entanglement \cite{BZ16}
and several other new sections.

\medskip

\section{Unitary operator bases and the Heisenberg groups}
\label{sec:uob}

\noindent Starting from a Hilbert space ${\cal H}$ of dimension $N$ we have another 
Hilbert space of dimension $N^2$ for free, namely the Hilbert-Schmidt space 
of all complex operators acting on ${\cal H}$, canonically 
isomorphic to the Hilbert space ${\cal H}\otimes {\cal H}^*$. It was introduced 
in Section 8.1 
and further explored in Chapter 9. 
Is it possible to find an orthonormal basis in ${\cal H}\otimes {\cal H}^*$ 
consisting solely of unitary operators? A priori this looks doubtful, since 
the set of unitary matrices has real dimension $N^2$, only one half the 
real dimension of ${\cal H}\otimes {\cal H}^*$. But physical observables 
are naturally associated to unitary operators, so if such bases exist they 
are likely to be important. They are called 
{\it unitary operator bases}, 
were introduced by Schwinger (1960) \cite{Schw60}, 
and heavily used by him \cite{Schw03}.

In fact unitary operator bases do exist, in great abundance.
 And we can ask for  more \cite{Kni96}. 
We can insist that the elements of the basis form a group. More 
precisely, let $\bar{G}$ be a finite group of order $N^2$, with identity element $e$. 
Let $U_g$ be unitary operators giving a projective representation of $\bar{G}$, 
such that 

\

1. $U_e$ is the identity matrix.

\smallskip

2. $g\neq e$ \hspace{3mm} $\Rightarrow$ \hspace{3mm} Tr$U_g = 0$. 

\smallskip

3. $U_gU_h = \lambda (g,h)U_{gh}$, where $|\lambda (g,h)| = 1$. 

\

\noindent (So $\lambda$ is a phase factor.) Then this collection of unitary 
matrices is a unitary operator basis. 
To see this, observe that
\begin{equation} U_g^\dagger = \lambda (g^{-1},g)^{-1}U_{g^{-1}} \ . \end{equation}

\noindent It follows that 
\begin{equation} g^{-1}h \neq e \hspace{5mm} \Rightarrow \hspace{5mm} 
\mbox{Tr}U_g^\dagger U_h = 0 \ , \end{equation}

\noindent and moreover that Tr$U_g^\dagger U_g = \mbox{Tr}U_e = N$. Hence these 
matrices are orthogonal with respect to the Hilbert-Schmidt inner product from 
Section 8.1. 
Unitary operator bases arising from a group in this way 
are known as unitary operator bases {\it of group 
type}, or as {\it nice error bases}---a name that comes from the theory of 
quantum computation (where they are used to discretize errors, thus making the 
latter correctable---as we will see in Section 17.7).

The question of the existence of nice error bases is a question in group theory. 
First of all we note that there are two groups involved in the construction, 
the group $G$ which is faithfully represented by the above formulas, and the 
{\it collineation group} $\bar{G}$ which is the group $G$ with all phase 
factors ignored. 
\index{group!collineation}
The group $\bar{G}$ is 
also known, in this context, as the {\it index group}. 
\index{group!index}
Unless $N$ is a prime number (in which case the nice error bases are 
essentially unique), there is a long list of possible index groups. An 
abelian index group is necessarily of the form $H\times H$, where $H$ 
is an abelian group. Non-abelian index groups are more difficult to 
classify, but it is known that every index group must be 
soluble. The classification problem has been studied by 
Klappenecker and R\"otteler \cite{Kla02}, making use of the 
classification of finite groups. They also maintain an on-line catalogue.
{\it Soluble groups} will reappear in Section \ref{sec:SICs}; 
for the moment let us just mention that all abelian groups are soluble.  

The paradigmatic example of a group $G$ giving rise to a unitary operator basis
is the {\it Weyl--Heisenberg group} $H(N)$.
This group appeared in many different 
contexts, starting in nineteenth century algebraic geometry, and in the beginnings 
of matrix theory \cite{Syl82}. In the twentieth century it took on a major role 
in the theory of elliptic curves \cite{Mum83}. Weyl (1932) \cite{Wey32} studied 
its unitary representations in his book on quantum mechanics. 
 The group $H(N)$ can be presented as follows.
 Introduce three group elements $X$, $Z$, and $\omega$. 
Declare them to be of order $N$: 
\begin{equation} X^N = Z^N = \omega^N = {\mathbbm 1} \ . \end{equation}

\noindent Insist that $\omega$ belongs to the centre of the group (it 
commutes with everything): 
\begin{equation} \omega X = X \omega \ , \hspace{8mm} \omega Z = Z\omega \ , \end{equation}

\noindent Then we impose one further key relation:
\begin{equation} ZXZ^{-1}X^{-1} = \omega \hspace{5mm} \Leftrightarrow \hspace{5mm} 
ZX = \omega XZ \ . \label{WH} \end{equation}

\noindent The Weyl--Heisenberg group consists of all `words' that can be written down 
using the three `letters' $\omega$, $X$, $Z$, subject to the above relations. It requires 
no great effort to see that it suffices to consider $N^3$ words of the form 
$\omega^tX^rZ^s$, where $t,r,s$ are integers modulo $N$. 

The Weyl--Heisenberg group admits an essentially unique unitary representation in 
dimension $N$. First we represent $\omega$ as multiplication with a phase factor  
which is a primitive root of unity, conveniently chosen to be 
\begin{equation} \omega = e^{2\pi i/N} \ . \end{equation}

\noindent If we further insist that $Z$ be represented by a diagonal operator we 
are led to the {\it clock-and-shift} representation
\begin{equation} Z|i\rangle = \omega^i|i\rangle \ , \hspace{6mm} X|i\rangle 
= |i+1\rangle \ . \end{equation}

\noindent The basis kets are labelled by integers modulo $N$. A very important area 
of application for the Weyl--Heisenberg group is that of time-frequency analysis of 
signals; then the operators $X$ and $Z$ may represent time delays and Doppler 
shifts of a returning radar wave form. But here we stick to the language of quantum 
mechanics and  refer to Howard et al. \cite{HCM06} for an introduction to 
signal processing and radar applications.

To orient ourselves we first write down the matrix form of the generators for $N = 3$, 
which is a good choice for illustrative purposes: 
\begin{equation} Z = \left( \begin{array}{ccc} 1 & 0 & 0 \\ 0 & \omega & 0 \\ 
0 & 0 & \omega^2 \end{array} \right) \ ,  \hspace{6mm} X = \left( \begin{array}{ccc} 
0 &  0 & 1 \\ 1 & 0 & 0 \\ 0 & 1 & 0 \end{array} \right) \ . \end{equation}

\noindent In two dimensions $Z$ and $X$ become the Pauli matrices $\sigma_z$ and 
$\sigma_x$ respectively. 
 We note 
the resemblance between eq. (\ref{WH}) and a special case of the equation that 
defines the original Heisenberg group, eq. (6.4). This explains why Weyl 
took this finite group as a toy model of the latter. We also note that although 
the Weyl--Heisenberg group has order $N^3$ its collineation group---the group modulo 
phase factors, which is the group acting on projective space---has order 
$N^2$. In fact the collineation group is an abelian product of two cyclic groups, 
$Z_N\times Z_N$. 
The slight departure from commutativity ensures an interesting representation theory.

There is a complication to notice at this point: because $\det{Z} = \det{X} = 
(-1)^{N+1}$, it matters if $N$ is odd or even. If $N$ is odd the Weyl--Heisenberg 
group is a subgroup of 
$SU(N)$, but if $N$ is even it is a subgroup of $U(N)$ only. Moreover, if $N$ is 
odd the $N$th power of every group element is the identity, but if $N$ is even 
we must go to the $2N$th power to say as much. (For $N = 2$ we find $X^2 = Z^2 
= {\mathbbm 1}$ but $(ZX)^2 = (i\sigma_y)^2 = - {\mathbbm 1}$.) These annoying 
facts make even dimensions significantly more difficult to handle, and leads to 
the definition
\begin{equation} \bar{N} = \left\{ \begin{array}{lll} N & \mbox{if} & N \ \mbox{is odd} \\ 
2N & \mbox{if} & N \ \mbox{is even.} \end{array} \right. \end{equation}

Keeping the complication in mind we turn to the problem of choosing suitable phase 
factors for the $N^2$ words that will make up our nice error basis. The peculiarities 
of even dimensions suggest an odd-looking move. 
We introduce the phase factor
\begin{equation} \tau \equiv - e^{\pi i/N} = - \sqrt{\omega} \ , 
\hspace{8mm} \tau^{\bar{N}} = 1 \ . \end{equation}

\noindent Note that $\tau$ is an $N$th root of unity only if $N$ is odd. 
Then we define the {\it displacement operators} 
\begin{equation} D_{\bf p} \equiv D_{r,s} = 
\tau^{rs}X^rZ^s \ , \hspace{8mm} 0 \leq r,s \leq \bar{N}-1 \ . \label{displace} \end{equation}

\noindent These are the words to use in the error basis. The double notation---using 
either $D_{\bf p}$ or $D_{r,s}$---emphasizes 
that it is convenient to view $r,s$ as the two components of a `vector' ${\bf p}$. 
Because of the phase factor $\tau$ the displacement operators are not actually 
in the Weyl--Heisenberg group, as originally defined, if $N$ is even. So there 
is a price to pay for this, but in return we get the group law in the form 
\begin{eqnarray}  D_{r,s}D_{u,v} = \omega^{us-vr}D_{u,v}D_{r,s} = 
\tau^{us-vr}D_{r+u,s+v} \nonumber \\ \Leftrightarrow \label{Hgrouplaw} \\ 
D_{\bf p}D_{\bf q} = \omega^{\Omega ({\bf p},{\bf q})}D_{\bf q}D_{\bf p} 
= \tau^{\Omega ({\bf p},{\bf q})}D_{{\bf p} + {\bf q}} 
\ .  \nonumber \end{eqnarray}

\noindent The expression in the exponent of the phase factors is anti-symmetric 
in the `vectors' that label the displacement operators. In fact $\Omega ({\bf p}, 
{\bf q})$ is a symplectic form (see Section 3.4), 
and a very nice object to encounter.

A desirable by-product of our conventions is 
\begin{equation} D^{-1}_{r,s} = D_{r,s}^\dagger = D_{-r,-s} \ . \end{equation}

\noindent Another nice feature is that the phase factor ensures that all displacement 
operators are of order $N$.  
On the other hand we have $D_{r+N,s} = \tau^{Ns}D_{r,s}$. This means that we have to live 
with a treacherous sign if $N$ is even, since the displacement operators 
are indexed by integers modulo $2N$ in that case. Even dimensions are unavoidably 
difficult to deal with. 
(We do not know who first commented that ``even dimensions 
are odd'', but he or she had a point.)

Finally, and importantly, we observe that 
\begin{equation} \mbox{Tr}D_{r,s} = \tau^{rs}\sum_{i=0}^{N-1}\omega^{si}\langle 
i|i+r\rangle = N\delta_{r,0}\delta_{s,0} \ . \end{equation}

\noindent Thus all the displacement operators except the identity are represented 
by traceless matrices, which means that the Weyl--Heisenberg group does indeed 
provide a unitary operator basis of group type, a nice error basis. Any complex 
operator $A$ on ${\cal H}_N$ can be written, uniquely, in the form 
\begin{equation} A = \sum_{r=0}^{N-1}\sum_{s=0}^{N-1}a_{rs}D_{r,s} \ , \label{WHUOB1} 
\end{equation}
\noindent where the expansion coefficients 
$a_{rs}$ are complex numbers given by 
\begin{equation} a_{rs} = \frac{1}{N}\mbox{Tr}D_{r,s}^\dagger A = 
\frac{1}{N}\mbox{Tr}D_{-r,-s}A \ . \label{WHUOB2} \end{equation}

\noindent Again we are using the Hilbert-Schmidt scalar product from Section 8.1.
Such expansions were called `quantum 
Fourier transformations' in Section 6.2. 
 
\section{Prime, composite, and prime power dimensions}
\label{sec:primepower}

The Weyl--Heisenberg group cares deeply whether the dimension $N$ is given by a 
prime number (denoted $p$), or by a composite number (say $N = p_1p_2$ or $N = p^K$). 
If $N = p$ then every element in the group has order $N$ (or $\bar{N}$ if $N = 2$), 
except of course for the identity element. If on the other hand $N = p_1p_2$ then 
$(Z^{p_1})^{p_2} = {\mathbbm 1}$, meaning that the element $Z^{p_1}$ has order $p_2$ 
only. This is a striking difference between prime and composite dimensions. 

The point is that we are performing arithmetic modulo $N$, which means that we regard 
all integers that differ by multiples of $N$ as identical. With this understanding, we 
can add, subtract, and multiply the integers freely. We say that integers modulo $N$ 
form a {\it ring}, just as the ordinary integers do.
\index{ring}
If $N$ is composite it can happen that $xy = 0$ modulo $N$, 
even though the integers $x$ 
and $y$ are non-zero. For instance, $2\cdot 2 = 0$ modulo 4. As Problem 
12.2 
should make clear, a ring is all we need to define 
a Heisenberg group, so we can proceed anyway. However, things work more smoothly if 
$N$ equals a prime number $p$, because of the striking fact that every non-zero integer 
has a multiplicative inverse modulo $p$. Integers modulo $p$ form a {\it field}, which 
by definition is a ring whose non-zero 
\index{field}
members form an abelian group under multiplication, so that we can perform division 
as well. In a field---the set of rational numbers is a standard example---we can 
perform addition, subtraction, multiplication, and division. In a ring---such as 
the set of all the integers---division sometimes fails. The distinction becomes 
important in Hilbert space once the latter is being organized by the Weyl--Heisenberg group.   

The field of integers modulo a prime $p$ is denoted ${\mathbb Z}_p$. When the 
dimension $N = p$ the 
operators $D_{\bf p}$ can be regarded as indexed by elements ${\bf p}$ of a two 
dimensional vector space. (We use the same notation for arbitrary $N$, but in 
general we need quotation marks around the word `vector'. In a true vector space 
the scalar numbers must belong to a field.) Note that this vector space contains $p^2$ 
vectors only. Now, whenever we encountered a vector space 
in this book, we tended to focus on the set of lines through its origin. This 
is a fruitful thing to do here as well. Each such line consists of all vectors 
${\bf p}$ obeying the equation ${\bf a}\cdot {\bf p} = 0$ for some fixed vector 
${\bf a}$. Since ${\bf a}$ is determined only up to an overall factor we obtain 
$p+1$ lines in all, given by 
\begin{equation} {\bf a} = \left( \begin{array}{c} a_1 \\ a_2 \end{array} \right) 
\in \left\{ \left( \begin{array}{c} 0 \\ 1 \end{array} \right) , 
\ \left( \begin{array}{c} 1 \\ 0 \end{array} \right) , \ \left( \begin{array}{c} 
1 \\ 1 \end{array} \right) , \ \dots \ , \ \left( \begin{array}{c} 1 \\ p-1 \end{array} 
\right) \right\} \ . \label{cyclsub} \end{equation}

\noindent This set of lines through the origin is a projective space with only 
$p+1$ points. Of more immediate interest is the fact that these lines through 
the origin correspond to cyclic subgroups of the Weyl--Heisenberg group, and 
indeed to its maximal abelian subgroups. (Choosing ${\bf a} = (1,x)$ gives the 
cyclic subgroup generated by $D_{-x,1}$.)  
The joint eigenbases of such subgroups are related in an interesting 
way, which will be the subject of Section \ref{sec:MUBs}. 

Readers who want a simple story are advised to ignore everything we say about 
non-prime dimensions. With this warning, we ask: 
What happens when the dimension is not prime? On the physical side this 
is often the case: we may build a Hilbert space of high dimension by taking 
the tensor product of a large number of Hilbert spaces which individually 
have a small dimension (perhaps to have a Hilbert space suitable for describing 
many atoms). This immediately suggests that it might be interesting to study the 
direct product $H(N_1)\times H(N_2)$ of two Weyl--Heisenberg groups, acting 
on the tensor product space ${\cal H}_{N_1}\otimes {\cal H}_{N_2}$. (Irreducible 
representations of a direct product of groups always act on the tensor product of 
their representation spaces.) Does this give something new?

On the group theoretical side it is known that the cyclic groups $Z_{N_1N_2}$ and 
$Z_{N_1}\times Z_{N_2}$ are isomorphic if and only if the integers ${N_1}$ and ${N_2}$ 
are relatively prime, that is to say if they do not contain any common factor. To see 
why, look at the examples $N = 2\cdot 2$ and $N = 2\cdot 3$. Clearly $Z_2\times Z_2$ 
contains only elements of order 2, hence it cannot be isomorphic to $Z_4$. On the other 
hand it is easy to verify that $Z_2\times Z_3 = Z_6$. This observation carries over 
to the Weyl--Heisenberg group: the groups $H(N_1N_2)$ and $H(N_1)\times H(N_2)$ 
are isomorphic if and only if $N_1$ and $N_2$ are relatively prime. Thus, in many 
composite dimensions including the physically interesting case $N = p^K$ we have a 
choice of more than one Heisenberg group to play with. They all form nice error bases. 
In applications to signal processing one sticks to $H(N)$ also when 
$N$ is large and composite, but in many-body physics and in the theory of quantum 
computation---carried out on tensor products of $K$ qubits, say---it is the 
{\it multipartite Heisenberg group} 
\index{group!multipartite Heisenberg}
$H(p)^{\otimes K} = H(p)\times H(p)\times \dots \times H(p)$ that comes to the fore.

There is a way of looking at the group $H(p)^{\otimes K}$ which is quite analogous to 
our way of looking at $H(p)$. It employs finite fields with $p^K$ elements, and 
de-emphasizes the tensor product structure of the representation space---which in 
some situations is a disadvantage, but it is worth explaining anyway, especially since 
\index{field!finite}
we will be touching on the theory of fields in Section \ref{sec:SICs}.
We begin by recalling how the field of complex numbers is 
constructed. One starts from the real field ${\mathbbm R}$, now called the 
{\it ground field}, and observes that the polynomial equation $P(x) \equiv x^2+1 = 0$ 
does not have a real solution. To remedy this the number ${\rm i}$ is introduced as a 
root of the equation $P(x) = 0$. With 
this new number in hand the complex number field ${\mathbbm C}$ is constructed as a 
two-dimensional vector space over ${\mathbbm R}$, with $1$ and ${\rm i}$ as basis vectors. 
To multiply two complex numbers together we calculate modulo the polynomial $P(x)$, 
which simply amounts to setting ${\rm i}^2$ equal to $-1$ whenever it occurs. The finite 
fields $GF(p^K)$ are defined in a similar way using the finite field ${\mathbbm Z}_p$ 
as a ground field. (Here `G' stands for Galois. For 
a lively introduction to finite fields we refer to Arnold \cite{Arn11}. For quantum 
mechanical applications we recommend the review by Vourdas \cite{Vou04}).

For an example, we may choose $p = 2$. The polynomial $P(x) = x^2+x+1$ 
has no zeros in the ground field ${\mathbbm Z}_2$, so we introduce an `imaginary' 
number $\alpha$ which is declared to obey $P(\alpha ) = 0$. Adding and multiplying in all 
possible ways, and noting that $\alpha^2 = \alpha +1$ (using binary 
arithmetic), we obtain a larger field having 
$2^2$ elements of the form $x_1+x_2\alpha$, where $x_1$ and $x_2$ are integers modulo 2. 
This is the finite field $GF(p^2)$. Interestingly its three non-zero elements can also 
be described as $\alpha$, $\alpha^2 = \alpha + 1$, and $\alpha^3 = 1$. The first 
representation is convenient for addition, the second for multiplication. We have a 
third description as the set of binary sequences $\{ 00, 01, 10, 11\}$, and consequently 
a way of adding and multiplying binary sequences together. This is very useful in the 
theory of error-correcting codes \cite{Pl82}. But this is by the way.

By pursuing this idea it has been proven that there exists a finite field $GF(N)$ 
with $N$ elements if and only if $N=p^K$ is a power of prime number 
$p$, and moreover that this finite field is unique for a given $N$ (which is by no means 
obvious because there will typically exist several polynomials of the same degree 
having no solution modulo $p$). These fields can be regarded as $K$ dimensional vector 
spaces over the ground field ${\mathbbm Z}_p$, which is useful when we do addition. 
When we do multiplication it is helpful to observe the (non-obvious) fact that 
finite fields always contain a {\it primitive element} in terms of which every 
non-zero element of the field can be written as the primitive 
element raised to some integer power, so the non-zero elements form a cyclic group. Some 
further salient facts are: 

\begin{itemize}

\item{Every element obeys $x^{p^K} = x$.}

\item{Every non-zero element obeys $x^{p^K-1} = 1$.}

\item{$GF(p^{K_1})$ is a subfield of $GF(p^{K_2})$ if and only if $K_1$ divides $K_2$.}

\end{itemize}

\noindent The field with $2^3$ elements is presented in Table \ref{tab:GF(8)}, 
and the field with $3^2$ elements is given as Problem 12.4. 

\begin{table}[h]
\caption{{\rm The field with 8 elements has irreducible polynomials} 
$P_1(x) = x^3+x+1$ {\rm and} 
$P_2(x) = x^3+x^2+1$; $\alpha$ {\rm is a root of} $P_1(x)$.}
  \smallskip
\hskip -0.3cm
{\renewcommand{\arraystretch}{1}
\begin{tabular}{|l|l|l|l|r|} \hline
Element & Polynomial  & tr$x$ & tr$x^2$ & order  \\ \hline 
0 = 0 & $x$ & 0 & 0 & --- \\ 
$\alpha^7 = 1$ & $x+1$ & 1 & 1 & 1 \\ 
$\alpha = \alpha$ & $P_1(x)$ & 0 & 0 & 7 \\ 
$\alpha^2 = \alpha^2$ & $P_1(x)$ & 0 & 0 & 7 \\
$\alpha^3 = \alpha + 1$ & $P_2(x)$ & 1 & 1 & 7 \\ 
$\alpha^4 = \alpha^2+\alpha $ & $P_1(x)$ & 0 & 0 & 7 \\ 
$\alpha^5 = \alpha^2+\alpha + 1$ & $P_2(x)$ & 1 & 1 & 7 \\ 
$\alpha^6 = \alpha^2+1$ & $P_2(x)$ & 1 & 1 & 7 \\
\hline \end{tabular}
}
\label{tab:GF(8)}
\end{table}

By definition the field theoretic {\it trace} is 
\begin{equation} \mbox{tr} \ x = x + x^p + x^{p^2} + \dots + x^{p^{K-1}} \ . \end{equation}

\noindent If $x$ belongs to the finite field ${\mathbbm F}_{p^K}$ its trace belongs to 
the ground field ${\mathbbm Z}_p$. Like the trace of a matrix, the field theoretic 
trace enjoys the properties that tr$(x+y) = \mbox{tr} x + \mbox{tr} y$ and tr$ax = 
a\mbox{tr} x$ for any integer $a$ modulo $p$. It is used to define the concept of 
{\it dual bases} for the field. A basis is simply a set of elements such that any 
element in the field can be expressed as a linear combination of this set, using 
coefficients in ${\mathbbm Z}_p$. Given a basis $e_i$ the dual basis $\tilde{e}_j$ is 
defined through the equation
\begin{equation} \mbox{tr}(e_i\tilde{e}_j) = \delta_{ij} \ . \end{equation}

\noindent For any field element $x$ we can then write, uniquely, 
\begin{equation} x = \sum_{i=1}^Kx_ie_i \hspace{5mm} \mbox{where} \hspace{5mm} 
x_i = \mbox{tr}(x\tilde{e}_i) \ . \end{equation}

\noindent From Table \ref{tab:GF(8)} we can deduce that the basis $(1,\alpha, \alpha^2)$ 
is dual to $(1,\alpha^2, \alpha)$, while the basis $(\alpha^3, \alpha^5, 
\alpha^6)$ is dual to itself.

Let us now apply what we have learned to the Heisenberg groups.
(For more details 
see Vourdas \cite{Vou04}, Gross \cite{Gro06}, and Appleby \cite{App09}).
Let $x, u,v$ be elements of the finite field $GF(p^K)$. Introduce 
an orthonormal basis and label its vectors by the field elements, 
\begin{equation} |x\rangle \ : \hspace{8mm} |0\rangle , \ |1\rangle , 
\ |\alpha \rangle , \ \dots , 
|\alpha^{p^K -2}\rangle \ . \end{equation}

\noindent Here $\alpha$ is a primitive element of the field, so there are $p^K$ 
basis vectors altogether. Using this basis we define the operators $X_u$, $Z_u$ by 
\begin{equation} X_u|x\rangle = |x+u\rangle \ , \hspace{5mm} Z_u|x\rangle 
= \omega^{\rm{tr}(xu)}|x\rangle \ , \hspace{5mm} \omega = e^{2\pi i/p} \ . 
\end{equation}

\noindent Note that $X_u$ is not 
equal to $X$ raised to the power $u$---this would 
make no sense, while the present definition does. In particular the phase 
factor $\omega$ is raised to an exponent that is just an ordinary integer modulo 
$p$. Due to the linearity of the field trace it is easily checked that 
\begin{equation} Z_uX_v = \omega^{{\rm tr}(vu)}X_vZ_u \ . \end{equation}

\noindent Note that it can happen that $X$ and $Z$ commute---it does happen for $GF(2^2)$, 
for which tr$(1) = 0$---so the definition takes 
some getting used to.

We can go on to define displacement operators 
\begin{equation} D_{\bf u} = \tau^{{\rm tr}u_1u_2}X_{u_1}Z_{u_2} \ , \hspace{6mm} 
\tau = -e^{i\pi/p} \ , \hspace{5mm} {\bf u} = \left( \begin{array}{c} u_1 \\ 
u_2 \end{array} \right) \ . \end{equation}

\noindent The phase factor has been chosen so that we obtain the desirable properties 
\begin{equation} D_{\bf u}D_{\bf v} = \tau^{\langle {\bf u},{\bf v}\rangle }
D_{{\bf u}+{\bf v}} \ , \hspace{8mm} D_{\bf u}D_{\bf v} = 
\omega^{\langle {\bf u},{\bf v}\rangle }D_{\bf v}D_{\bf u} 
\hspace{8mm} D_{\bf u}^\dagger = D_{- {\bf u}} \ . \end{equation}

\noindent Here we introduced the symplectic form 
\begin{equation} \langle {\bf u}, {\bf v}\rangle = {\rm tr}(u_2v_1 - u_1v_2) \ . 
\end{equation}

\noindent So the formulas are arranged in parallel with those used to describe $H(N)$. 
It remains to show that the resulting group is isomorphic to the one obtained by 
taking $K$-fold products of the group $H(p)$. 

There do exist isomorphisms between the two groups, but there does not exist a 
canonical isomorphism. Instead we begin by choosing a pair of dual bases for the field, 
obeying tr$(e_i\tilde{e}_j) = \delta_{ij}$. We can then expand a given element of the 
field in two different ways, 
\begin{equation} x = \sum_{i=1}^Kx_ie_i = \sum_{r=1}^K\tilde{x}_i\tilde{e}_i \hspace{5mm} 
\Leftrightarrow \hspace{5mm} \left\{ \begin{array}{l} x_i = {\rm tr}(x\tilde{e}_i) \\ 
\tilde{x}_i = {\rm tr}(xe_i) \ . \end{array} \right. \end{equation} 

\noindent We then introduce an isomorphism between ${\cal H}_{p^K}$ and ${\cal H}_p\otimes 
\dots \otimes {\cal H}_p$, 
\begin{equation} S|x\rangle = |x_1\rangle \otimes |x_2\rangle \otimes \dots \otimes 
|x_K\rangle \ . \end{equation}

\noindent In each $p$ dimensional factor space we have the group 
$H(p)$ and the displacement operators 
\begin{equation} D^{(p)}_{i,j} = \tau^{rs}X^rZ^s \ , \hspace{8mm} r,s \in {\mathbb Z}_p \ . 
\end{equation}

\noindent To set up an isomorphism between the two groups we expand 
\begin{equation} {\bf u} = \left( \begin{array}{c} u_1 \\ u_2 \end{array} \right) = 
\left( \begin{array}{c}\sum_iu_{1i}e_i \\ \sum_i\tilde{u}_{2i}\tilde{e}_i 
\end{array} \right) \ . \end{equation}

\noindent Then the isomorphism is given by 
\begin{equation} D_{\bf u} = S^{-1}\left( D^{(p)}_{u_{11},\tilde{u}_{21}} \otimes 
D^{(p)}_{u_{12},\tilde{u}_{22}}\otimes \dots \otimes D^{(p)}_{u_{1K},\tilde{u}_{2K}} 
\right) S \ . \end{equation}

\noindent The verification consists of a straightforward calculation showing that 
\begin{equation} D_{\bf u}D_{\bf v} = \tau^{\sum_i(\tilde{u}_{2i}v_{1i} - u_{1i}
\tilde{v}_{2i})} D_{{\bf u} + {\bf v}} = \tau^{\langle u,v\rangle }D_{{\bf u} + {\bf v}} 
\ . \end{equation}
 
\noindent It must of course be kept in mind that the isomorphism inherits the 
arbitrariness involved in choosing a field basis. Nevertheless this  
reformulation has its uses, notably because we can again regard the set of displacement 
operators as a vector space over a field, and we can obtain $N+1 = p^K + 1$ maximal 
abelian subgroups from the set of lines through its origin. However, unlike in the 
prime dimensional case, we do not obtain every maximal abelian subgroup from this 
construction \cite{KRBSS09}. 

\section{More unitary operator bases}
\label{sec:Werner}

Do all interesting things come from groups? For unitary operator bases the answer 
is a resounding `no'. We begin with a slight reformulation of the problem. Instead 
of looking for special bases in the ket/bra Hilbert space ${\cal H}\otimes 
{\cal H}^*$ we look for them in the ket/ket Hilbert space 
${\cal H}\otimes {\cal H}$. We relate the two spaces with a map that 
interchanges their computational bases, $|i\rangle \langle j| \leftrightarrow 
|i\rangle |j\rangle$, while leaving the components of the vectors unchanged. 
A unitary operator $U$ with matrix elements $U_{ij}$ then corresponds to the state
\begin{equation} |U\rangle = \frac{1}{\sqrt{N}}
\sum_{i,j = 0}^{N-1}U_{ij}|i\rangle |j\rangle \ \in \ {\cal H}\otimes {\cal H} \ . 
 \label{ustate} \end{equation}

\noindent States of this form are said to be maximally entangled, and we will return 
to discuss them in detail in Section 16.3.
A special example, obtained 
by setting $U_{ij} = \delta_{ij}$, appeared already in eq. (11.21).
For now we just observe that the task of finding a unitary operator basis for 
${\cal H}\otimes {\cal H}^*$ is equivalent to that of finding a maximally 
entangled basis for ${\cal H}\otimes {\cal H}$.    

A rich supply of maximally entangled bases can be obtained using two concepts 
imported from discrete mathematics: Latin squares and (complex) Hadamard matrices. 
We explain the construction for $N = 3$, starting with the special state  
\begin{equation} |\Omega \rangle = \frac{1}{\sqrt{3}}(|0\rangle |0\rangle + 
|1\rangle |1\rangle + |2\rangle |2\rangle ) \ . \end{equation}

\noindent Now we bring in a {\it Latin square}. 
By definition this is an array of 
$N$ columns and $N$ rows containing a symbol from an alphabet of $N$ letters in 
each position, subject to the restriction that no symbol occurs twice in any row or 
in any column. The study of Latin squares goes back to Euler; Stinson 
\cite{Sti04} provides a good account. If the reader has spent time on sudokos she 
has worked within this tradition already. Serious applications of Latin squares, 
to randomization of agricultural experiments, 
were promoted by Fisher \cite{Fis35}.
 
We use a Latin square to expand our maximally 
entangled state into $N$ orthonormal maximally entangled states. An example with 
$N = 3$ makes the idea transparent:

\begin{equation} \begin{array}{|c|c|c|} \hline 0 & 1 & 2 \\ \hline 1 & 2 & 0 \\ \hline 
2 & 0 & 1 \\ \hline \end{array} \hspace{4mm} \rightarrow \hspace{4mm} 
\begin{array}{l} 
|\Omega_0 \rangle = \frac{1}{\sqrt{3}}(|0\rangle |0\rangle + |1\rangle |1\rangle + 
|2\rangle |2\rangle ) \\ \\ 
|\Omega_1 \rangle = \frac{1}{\sqrt{3}}(|0\rangle |1\rangle + 
|1\rangle |2\rangle + |2\rangle |0\rangle ) \\ \\ 
|\Omega_2 \rangle = \frac{1}{\sqrt{3}}(|0\rangle |2\rangle + 
|1\rangle |0\rangle + |2\rangle |1\rangle ) \ . \end{array}
\label{maxentbasprel} \end{equation}

\noindent The fact that the three states (in ${\cal H}_9$) are mutually orthogonal is an 
automatic consequence of the properties of the Latin square. But we want $3^2$ 
orthonormal states. To achieve this we bring in a {\it complex Hadamard matrix}, 
that is to say a unitary matrix each of whose elements have the same 
modulus. The {\it Fourier matrix} $F$, whose matrix 
elements are 
\begin{equation} F_{jk} = \frac{1}{\sqrt{N}}\omega^{jk} = 
\frac{1}{\sqrt{N}}(e^{\frac{2\pi i}{N}})^{jk} 
\ , \hspace{8mm} 0 \leq j,k \leq N - 1 \ . \label{Fourier}\end{equation}  

\noindent provides an example that works for every $N$. For $N = 3$ 
it is an essentially unique example.
For complex Hadamard matrices in general, see Tadej and \.Zyczkowski \cite{TaZ06}, and Sz{\"o}ll{\H{o}}si \cite{Szo11}.
 We use such a matrix to 
expand the vector $|\Omega_0\rangle$ according to the pattern 

\begin{equation} \frac{1}{\sqrt{3}}\left[ \begin{array}{ccc} 1 & 1 & 1 \\ 1 & \omega & \omega^2 \\ 
1 & \omega^2 & \omega \end{array} \right] \hspace{2mm} \rightarrow \hspace{2mm} 
\begin{array}{l} 
|\Omega_{00} \rangle = \frac{1}{\sqrt{3}}(|0\rangle |0\rangle + \ \ |1\rangle |1\rangle + 
\ \ |2\rangle |2\rangle ) \\ \\ 
|\Omega_{01} \rangle = \frac{1}{\sqrt{3}}(|0\rangle |0\rangle + 
\omega |1\rangle |1\rangle + \omega^2|2\rangle |2\rangle ) \\ \\ 
|\Omega_{02} \rangle = \frac{1}{\sqrt{3}}(|0\rangle |0\rangle + 
\omega^2|1\rangle |1\rangle + \omega|2\rangle |2\rangle ) \ . \end{array}
\end{equation}

\noindent The orthonormality of these states is guaranteed by the properties of  
the Hadamard matrix, and they are obviously maximally entangled. Repeating the 
construction for the remaining states in (\ref{maxentbasprel}) yields a full orthonormal 
basis of maximally entangled states. In fact for $N = 3$ we obtained nothing new; we 
simply reconstructed the unitary operator basis provided by the Weyl--Heisenberg 
group. The same is true for $N = 2$, where the analogous basis is known as the 
Bell basis, and will reappear in eq. (16.1). 

The generalization to any $N$ should be clear, especially if we formalize the 
notion of Latin squares a little. This will also provide some clues how the set 
of all Latin squares can be classified. First of all Latin squares exist for any 
$N$, because the multiplication table of a finite group is a Latin square. But most 
Latin squares do not arise in this way. So how many Latin squares are there? 
To count them one may first agree to present them in reduced form, which means 
that the symbols appear in lexicographical order in the first row and the first 
column. This can always be arranged by permutations of rows and columns. But there 
are further natural equivalences in the problem. A Latin square can be presented as 
$N^2$ triples $(r,c,s)$, for `row, column, and symbol'. The rule is that in this 
collection all pairs $(r,c)$ are different, and so are all pairs $(r,s)$ and 
$(s,c)$. So we have $N$ non--attacking rooks on a cubic chess board of size $N^3$. 
In this view the symbols are on the same footing as the rows and columns, and can 
be permuted. A formal way of saying this is to introduce a map $\lambda \ : \ 
{\mathbbm Z}_N \times {\mathbbm Z}_N \rightarrow {\mathbbm Z}_N$, where ${\mathbbm Z}_N$ denotes 
the integers modulo $N$, such that the maps
\begin{equation} i \rightarrow \lambda (i,j) \ , \hspace{8mm} i \rightarrow 
\lambda (j,i) \ , \end{equation}

\noindent are injective for all values of $j$. Two Latin squares are said to be 
{\it isotopic} if they can be related by permutations within the three copies of 
${\mathbbm Z}_N$ involved in the map. The classification of Latin squares under 
these equivalences was completed up to $N = 6$ by Fisher and his collaborators \cite{Fis35}, 
but for higher $N$ the numbers grow astronomical. See Table \ref{tab:latin}. 

\begin{table}
\caption{ Counting Latin squares and equivalence classes of complex Hadamard 
matrices of order $N$. The result for Hadamard matrices of size $N = 6$ 
is quite recent \cite{Szo12, BoZh15}, for $N = 7$ we give a lower bound only, 
and little is known beyond that. For $N = 8$ there are 1676257 isotopy 
classes of Latin squares, so in both cases we have good reasons to break 
off the table.
}
  \smallskip 
\hskip -0.3cm
{\renewcommand{\arraystretch}{1.67}
\begin{tabular}
[c]{|l|| r|r|r||r|}\hline \hline 
$N$ & Latin squares & Reduced squares  & Isotopic squares & Hadamards \\
 \hline
2 & 2
         & 1 & 1 & 1 \\
3 & 12
         & 1 & 1 & 1 \\
4 & 576
          & 4 &  2 & $\infty$ \\
5  & 161280
          & 56 & 2 & 1 \\
6 & 812851200
         & 9408 & 22 & $\infty^4 + 1$ \\			
7 & 61479419904000 & 16942080 & 564 & $\geq \infty + 1$ \\						
\hline \hline
\end{tabular}
}
\label{tab:latin}
\end{table}

The second ingredient in the construction, complex Hadamard matrices, also raises 
a difficult classification problem. 
The appropriate equivalence relation for this 
classification includes permutation of rows and columns, as well as acting with 
diagonal unitaries from the left and from the right. Thus we adopt the equivalence relation
\begin{equation} H' \sim PDHD'P' \ , \label{Uffe} \end{equation}

\noindent where $D,D'$ are diagonal unitaries and $P,P'$ are permutation matrices. 
For $N = 2,3$, and $5$, all complex Hadamard matrices are equivalent to the Fourier matrix. 
For $N = 4$ there exists a one-parameter family of inequivalent examples, including a purely real Hadamard matrix --
 see Table \ref{tab:latin}, and also Problem 12.3.

The  freely adjustable phase factor was discovered by Hadamard \cite{Had93}. 
It was adjusted in experiments performed many years later \cite{Lai12}. 
Karlsson wrote down a three parameter family of $N = 6$ complex Hadamard matrices 
in fully explicit and remarkably elegant form \cite{Kar11}. Karlsson's family is qualitatively more interesting than the $N = 4$ example.

Real Hadamard matrices (having entries $\pm 1$) can exist only if $N = 2$ or 
$N = 4k$. Paley conjectured \cite{Pal33} that in 
these cases they always exist, but his conjecture remains open.
The smallest unsolved 
case is $N = 668$. Real Hadamard matrices have many uses, and discrete 
mathematicians have spent much effort constructing them \cite{Hor07}.

With these ingredients in hand we can write down the vectors in a maximally 
entangled basis as
\begin{equation} |\Omega_{ij}\rangle = \frac{1}{\sqrt{N}}\sum_{k=0}^{N-1}
H_{jk}|k\rangle |\lambda (i,k)\rangle \ , \end{equation}

\noindent where $H_{ik}$ are the entries in a complex Hadamard matrix and 
the function $\lambda$ defines a Latin square. 
(A quantum variation on the 
theme of Latin squares, giving an even richer supply, is known 
\cite{MV15}).
The construction above is due to Werner \cite{Wer01}. 
Since it relies on arbitrary Latin squares 
and arbitrary complex Hadamard matrices we get an enormous supply 
of unitary operator bases out of it.

 This many groups do not exist, so most of these bases cannot 
be obtained from group theory. Still some nice error bases---in 
particular, the ones coming from the Weyl-Heisenberg group---do 
turn up (as, in fact, it did in our $N = 3$ example). The converse 
question, whether all nice error bases come from Werner's construction, has an 
answer, namely `no'. The examples constructed in this section are all {\it monomial}, 
meaning that the unitary operators can be represented by matrices having only one 
non-zero entry in each row and in each column. Nice error bases not of 
this form are known. Still it is interesting to observe that the operators 
in a nice error basis can be represented by quite sparse matrices---they 
always admit a representation in which at least one half of the matrix elements 
equal zero \cite{Kla05}.

\section{Mutually unbiased bases}
\label{sec:MUBs}

\noindent Two orthonormal bases $\{ |e_i\rangle \}_{i=0}^{N-1}$ and 
$\{ |f_j\rangle \}_{j=0}^{N-1}$ 
are said to be {\it complementary} or {\it unbiased} if 
\begin{equation} |\langle e_i|f_j\rangle |^2 = \frac{1}{N} \label{MUB1} 
\end{equation}

\noindent for all possible pairs of vectors consisting of one vector 
from each basis. If a system is prepared in a state belonging to one 
of the bases, then no information whatsoever is available about the 
outcome of a von Neumann measurement using the complementary basis. 
The corresponding observables are complementary in the sense of Niels 
Bohr, whose main concern was with the complementarity between position 
and momentum as embodied in the formula 
\begin{equation} |\langle x|p\rangle |^2 = \frac{1}{2\pi} \ . \end{equation}

\noindent The point is that the right hand side is a constant. Its actual value 
is determined by the probabilistic interpretation only when the dimension 
of Hilbert space is finite.

To see why a set of mutually unbiased bases may be a desirable thing to 
have, suppose we are given an unlimited supply of identically prepared 
$N$ level systems, and that we are asked to determine the $N^2-1$ 
parameters in the density matrix $\rho$. 
That is, we are asked to perform quantum state tomography on $\rho$. 
Performing the same von Neumann measurement on every copy will determine  
a probability vector with $N-1$ parameters. But 
\begin{equation} N^2-1 = (N+1)(N-1) \ . \label{cool}\end{equation}

\begin{figure}[ht]
        \centerline{ \hbox{
                 \epsfig{figure=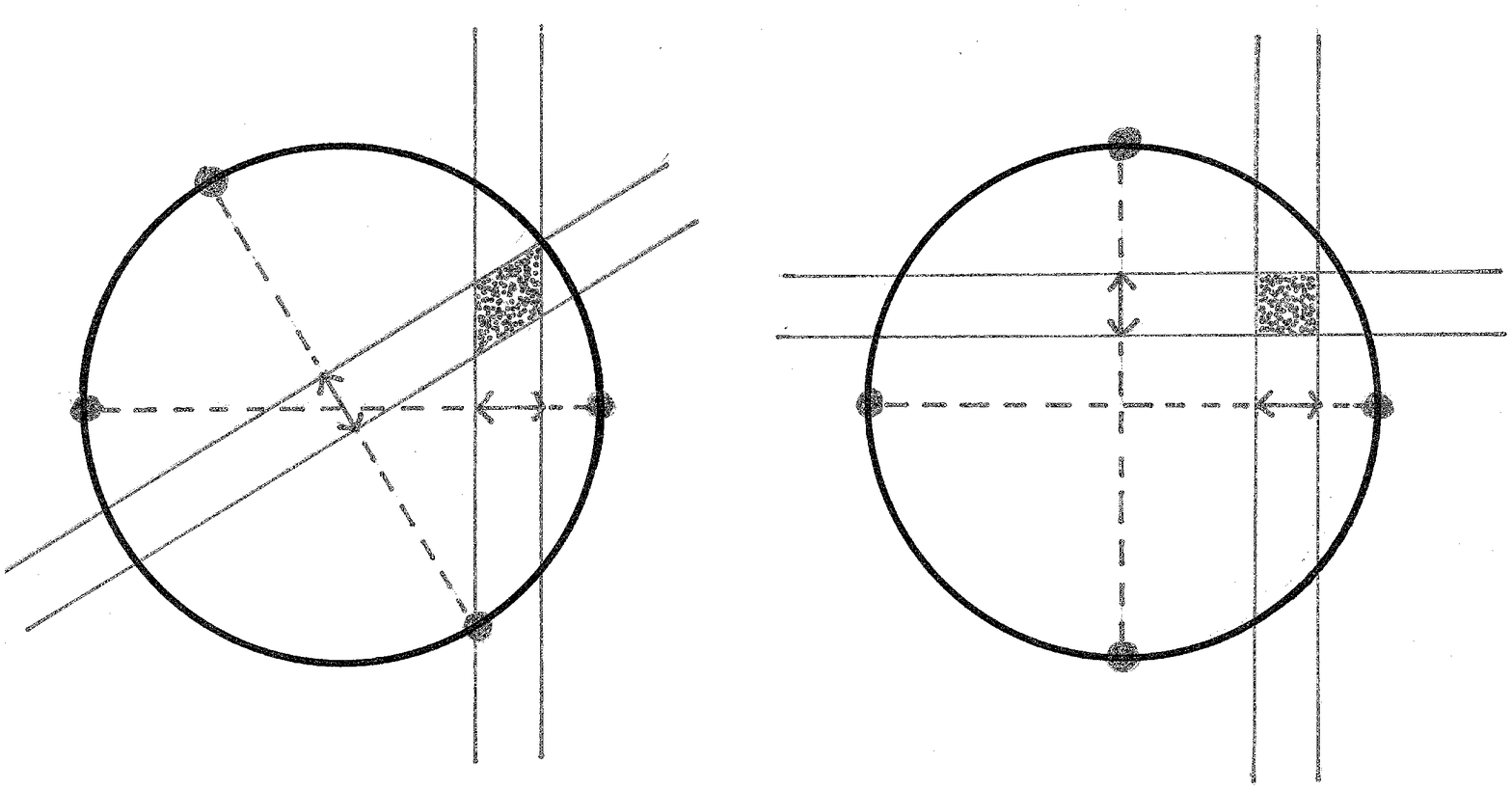, width=85mm}}}
        \caption{Quantum state tomography for a real qubit (a `rebit'), 
using pairs of bases. The statistics of each measurement determine the position of the density matrix only up to a statistical spread. 
Choosing the two bases representing the measurement to be unbiased---this is done for the rebit on the right---will minimize the resulting uncertainty 
(represented by the shaded area). 
Readers who recall Figure 2.11 
will realize that we suppress some complications here.}
        \label{fig:mub2}
\end{figure}

\noindent Hence we need to perform $N+1$ different measurements to fix $\rho$. 
If---as is likely to happen in practice---each measurement can be performed 
a finite number of times only, then there will be a statistical spread, 
and a corresponding uncertainty in the determination 
of $\rho$. Figure \ref{fig:mub2} is intended to suggest (correctly as it 
turns out) that the best result will be achieved if the $N+1$ measurements 
are performed using {\it Mutually Unbiased Bases} (abbreviated {\it MUB} 
from now on). 

MUB have numerous other applications, notably to quantum 
cryptography. The original BB84 protocol for quantum key distribution 
\cite{BB84} used a pair of qubit MUB. Going to larger sets of MUB in 
higher dimensions yields further advantages \cite{CBKG02}. The bottom line 
is that MUB are of interest when one is trying to find or hide 
information. Further applications 
include entanglement detection in the laboratory \cite{SHBAH12}, a famous 
retrodiction problem \cite{EA01,ReWe07}, and more.

\begin{figure}[htbp]
        \centerline{ \hbox{
                 \epsfig{figure=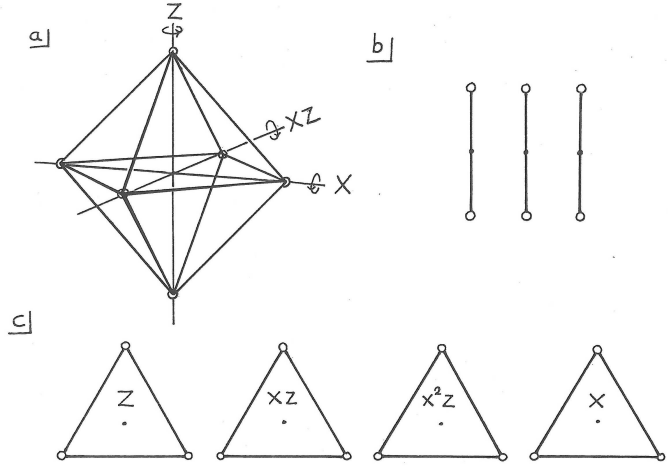, width=90mm}}}
        \caption{a) When $N = 2$ a complete set of MUB forms a regular octahedron 
inscribed in the Bloch sphere. Each of the bases is an eigenbasis for some element in 
the Weyl--Heisenberg group, represented in the figure by a $\pi$-rotation through an 
axis. b) A poor man's picture of the octahedron, shown as three 1-simplices, assumed 
to be mutually orthogonal and to intersect only at the maximally mixed state. c) A poor 
man's picture of four MUB in three dimensions---in the 8-dimensional Bloch 
space it consists of four 2-simplices spanning totally orthogonal subspaces, 
intersecting only at the maximally mixed state. Each basis is the eigenbasis 
of a cyclic subgroup generated by some element of the Weyl--Heisenberg group.}
        \label{fig:mub1}
\end{figure}

Our concern will be to find out how many MUB exist in a given dimension $N$. 
The answer will tell us something about the shape of the convex body of mixed 
quantum states. To see this, note that when $N = 2$ the pure states making up 
the three MUB form the corners of a regular octahedron inscribed in the 
Bloch sphere. See Figure \ref{fig:mub1}. 
Now let the dimension of Hilbert space be $N$. The set of mixed 
states has real dimension $N^2-1$, and it has the maximally mixed state as its 
natural origin. An orthonormal basis corresponds to a regular simplex 
$\Delta_{N-1}$ centred at the origin. It spans an $(N-1)$-dimensional plane 
through the origin. Using the trace inner product we find that 
\begin{equation} |\langle e_i|f_j\rangle |^2 = \frac{1}{N} \hspace{5mm} \Rightarrow 
\hspace{5mm} \mbox{Tr}\left( |e_i\rangle \langle e_i| - \frac{1}{N}{\mathbbm 1}\right) 
\left( |f_j\rangle \langle f_j| - \frac{1}{N}{\mathbbm 1}\right) = 0 \ . \end{equation}

\noindent Hence the condition that two such simplices represent a pair of MUB 
translates into the condition that the two $(N-1)$-planes be totally orthogonal, in 
the sense that every Bloch vector in one of them is orthogonal to every Bloch vector in 
the other. But the central equation of this section, namely (\ref{cool}), implies 
that there is room for at most $N+1$ totally orthogonal $(N-1)$-planes. 
It follows that there can exist at most $N+1$ MUB. But it does not 
at all follow that this many MUB actually exist. Our collection of $N+1$ 
simplices form an interesting convex polytope with $N(N+1)$ vertices, and what 
we are asking is whether this {\it complementarity polytope} can be inscribed
\index{polytope!complementarity} 
into the convex body ${\cal M}^{(N)}$ of density matrices. In fact, given our 
caricature of this body, as the stitching found on a tennis ball (Section 8.6), 
this does seem a little unlikely (unless $N = 2$). 

Anyway a set of $N+1$ MUB in $N$ dimensions is referred to as a {\it complete 
set}. Do such sets exist? If we think of a basis as given by the column vectors 
of a unitary matrix, and if the basis is to be unbiased relative to the 
computational basis, then that unitary matrix must be a complex Hadamard matrix. 
Classifying pairs of MUB is equivalent to classifying such matrices. In Section 
\ref{sec:Werner} we saw that they exist for all $N$, often in 
abundance. To be specific, let the identity matrix and the Fourier matrix 
(\ref{Fourier}) represent a pair of MUB. Can we find a third basis, 
unbiased with respect to both? 
Using $N = 3$ as an illustrative example we find that Figure 4.10 
gives the story away. A column vector in a complex Hadamard matrix corresponds 
to a point on the maximal torus in the octant picture. The twelve column vectors 
\begin{equation} \left[ \begin{array}{ccc|ccc|ccc|ccc} 1 & 0 & 0 & 1 & \omega^2 & 
\omega^2 & 1 & \omega & \omega & 1 & 1 & 1 \\ 
0 & 1 & 0 & \omega^2 & 1 & \omega^2 & \omega & 1 & \omega & 1 & \omega & \omega^2 \\ 
0 & 0 & 1 & \omega^2 & \omega^2 & 1 & \omega & \omega & 1 & 1 & \omega^2 & \omega 
\end{array} \right] \end{equation}

\noindent form four MUB, and it is clear from the picture that this is the 
largest number that can be found. (For convenience we did not normalize the 
vectors. The columns of the Fourier matrix $F$ were placed on the right.) 

We multiplied the vectors in the two bases in the middle with phase 
factors in a way that helps to make the pattern memorable. Actually there is 
a bit more to it. They form {\it circulant matrices} of size 3, meaning that
they can be obtained by cyclic permutations of their first row. Circulant 
matrices have the nice property that $F^\dagger CF$ is a diagonal matrix for 
every circulant matrix $C$.   

The picture so far is summarized in Figure \ref{fig:mub1}. The key observation 
is that each of the bases is an eigenbasis for a cyclic subgroup of the 
Weyl--Heisenberg group. As it turns out this generalizes straightforwardly 
to all dimensions such that $N = p$, where $p$ is an odd prime. We gave a 
list of $N+1$ cyclic subgroups in eq. 
(\ref{cyclsub}). Each cyclic subgroup consists of a complete set of commuting 
observables, and they determine an eigenbasis. We denote 
the $a$-th vector in the $x$-th eigenbasis as $|x,a\rangle$, 
and we have to solve 
\begin{equation} D_{0,1}|0,a\rangle = \omega^a|0,a\rangle \ , \hspace{3mm} 
D_{x,1}|x,a\rangle = \omega^a|x,a\rangle \ , \hspace{3mm} D_{-1,0}|\infty , a\rangle 
= \omega^a|\infty , a\rangle \ . \end{equation}

\noindent Here $x = 1, \dots , p-1$, but in the spirit of projective geometry we can 
extend the range to include $0$ and $\infty$ as well. The solution, with 
$\{ |e_r\rangle \}_{r=0}^{p-1}$ denoting the computational basis, is  
\begin{equation} |0,a\rangle = |e_a\rangle \ , \hspace{3mm} 
|x,a\rangle = \frac{1}{\sqrt{p}}\sum_{r=0}^{p-1}\omega^{\frac{(r-a)^2}{2x}} |e_r\rangle \ ,
\hspace{3mm} |\infty , a\rangle = \frac{1}{\sqrt{p}}\sum_{r=0}^{p-1}\omega^{ar}|e_r\rangle \ .  
\label{ivanovic} \end{equation} 

\noindent It is understood that if `1/2' occurs in an exponent it denotes the 
inverse of 2 in arithmetic modulo $p$ (and similarly for `1/x'). There are $p-1$ bases 
presented as columns of circulant matrices, and we use $\infty$ to label the Fourier 
basis for a reason (Problem 12.5) 
One can  show directly (as done in 1981  by Ivanovi\'c \cite{Iv81}, 
whose interest was in state tomography)
 that these bases form a complete set of MUB 
(Problem 12.6),
but a simple and remarkable theorem will save us from this effort.

Interestingly there is an alternative way to construct the complete set. 
When $N = 2$ it is clear that we can start with the eigenbasis of 
the group element $Z$ (say), choose a point on the equator, and 
then apply all the transformations effected by the Weyl--Heisenberg 
group. The resulting orbit will consist of $N^2$ points, and if the 
starting point is judiciously chosen they will form $N$ MUB, all of 
them unbiased to the eigenbasis of $Z$. Again the construction works 
in all prime dimensions $N = p$, and indeed the resulting complete set 
is equivalent to the previous one in the sense that there exists a 
unitary transformation taking the one to the other.
This 
construction is due to Alltop (1980), 
whose interest was in radar applications 
\cite{Al80}. Later references have made the construction more transparent 
and more general \cite{Bl14, KlaRoe04}.

The central theorem about MUB existence is due to 
Bandyopadhyay, Boykin, Roychowdhury, and Vatan \cite{BBRV02}.

\smallskip
\noindent {\bf The BBRV theorem}. {\sl A complete set of MUB exists 
in ${\mathbb C}^N$ if and only if there exists a unitary operator basis which 
can be divided into $N+1$ sets of $N$ commuting operators such that 
the sets have only the unit element in common.} 
\smallskip

\begin{figure}[ht]
        \centerline{ \hbox{
                 \epsfig{figure=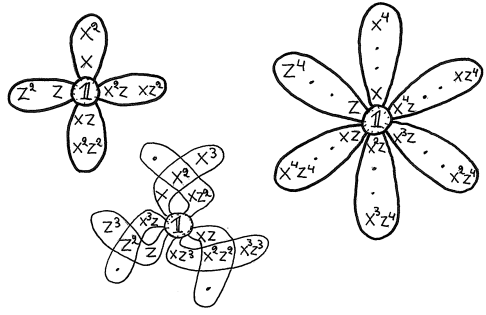, width=65mm}}}
        \caption{For $N = 3$ and $5$ the cyclic subgroups of the Weyl-Heisenberg
 group form a flower. For $N = 4$ the petals get in each other's way
 and the construction fails.}
        \label{fig:mub3}
\end{figure}

\noindent Let us refer to unitary operator bases of this type as {\it flowers}, 
and the sets into which they are divided as {\it petals}. Note that it is eq. (\ref{cool}) 
that makes flowers possible. There can be at most $N$ mutually commuting orthogonal 
unitaries since, once they are diagonalized, the vectors defined by their diagonals 
must form orthogonal vectors in $N$ dimensions. The Weyl--Heisenberg 
groups are flowers if and only if $N$ is a prime number---as exemplified 
in Figure \ref{fig:mub3}. So the fact that (\ref{ivanovic}) gives a complete 
set of $N+1$ MUB whenever $N$ is prime follows from the theorem. 

We prove the BBRV theorem one way. Suppose a complete set of MUB exists.  
We obtain a maximal set of commuting Hilbert-Schmidt orthogonal unitaries by carefully 
choosing $N$ unitary matrices $U_r$, with $r$ between $0$ and $N-1$:
\begin{equation} U_r = \sum_r\omega^{ri}|e_i\rangle \langle e_i| \hspace{5mm} 
\Rightarrow \hspace{5mm} \mbox{Tr}U_r^\dagger U_s = \sum_i\omega^{i(r-s)} = 
N\delta_{rs} \ . \end{equation}

\noindent If the bases $\{ |e_i\rangle \}_{i=0}^{N-1}$ and 
$\{ |f_i\rangle \}_{i=0}^{N-1}$ are unbiased we can form two such sets, 
and it is easy to check that 
\begin{equation} V_r = \sum_r\omega^{ri}|f_i\rangle \langle f_i| \hspace{3mm} 
\Rightarrow \hspace{3mm} \mbox{Tr}V_r^\dagger U_s = \frac{1}{N}\sum_{i,j}\omega^{-ir}
\omega^{js} = N\delta_{r,0}\delta_{s,0} \ . \end{equation} 

\noindent Hence $U_r \neq V_s$ unless $r = s = 0$. It may seem as if we 
have constructed two cyclic subgroups of the 
Weyl-Heisenberg group, but in fact we have not since we have said nothing about
the phase factors that enter into the scalar products $\langle e_i|f_j\rangle$. 
But this is as may be. It is still clear that if we go in this way, 
we will obtain a flower from a collection of $N+1$ MUB. Turning this into an `if 
and only if' statement is not very hard \cite{BBRV02}.

We have found flowers in all prime dimensions. What about $N = 4$? At this 
point we recall the Mermin square (5.61).
It defines (as it must, if one looks at the proof of the BBRV theorem) two distinct triplets 
of MUB. It turns out, however, that these are examples of {\it unextendible} 
sets of MUB, that cannot be completed to complete sets (a fact that can be 
ascertained without calculations, if the reader has the prerequisites needed 
to solve Problem 12.7. 
A more constructive observation is that the operators occurring 
in the square belong to the 
2--partite Heisenberg group $H(2)\times H(2)$. This group is also a unitary 
operator basis, and it contains no less than 
15 maximal abelian subgroups, or petals in our language. Denoting the elements 
of the collineation group of $H(2)$ as $X = \sigma_x, \ Y = \sigma_y, \ Z= \sigma_z$, 
and the elements of the 2--partite group as (say) $XY = X\otimes Y$, we label 
the petals as  
\begin{equation} \begin{array}{lll} 1 = \{ {\mathbbm 1}Z, Z{\mathbbm 1}, ZZ\} 
& 2 = \{ X{\mathbbm 1}, {\mathbbm 1} X, XX\}  & 3 = \{ XZ, ZX, YY \} \\
4 = \{ {\mathbbm 1}Z, X{\mathbbm 1}, XZ\}  & 5 = 
\{ Z{\mathbbm 1}, {\mathbbm 1}X, ZX\} & 6 = \{ ZZ, XX, YY \} \end{array}     
\label{Merminpetals} \end{equation}

\noindent (these are the petals occurring in the Mermin square), and
\begin{equation} \begin{array}{lll} \ 7 = \{ {\mathbbm 1}Z, Y{\mathbbm 1}, YZ\}  
& \ 8 = \{ Z{\mathbbm 1}, {\mathbbm 1}Y, ZY \}  & \ 9 = \{ X{\mathbbm 1}, {\mathbbm 1}Y, XY \} \\
10 = \{ {\mathbbm 1}X, Y{\mathbbm 1}, YX \}  & 11 = 
\{ {\mathbbm 1}Y, Y{\mathbbm 1}, YY \}  & 12 = \{ XY, YX, ZZ \} \\ 
13 = \{ XZ, YX, ZY \}  & 14 = 
\{ XY, YZ, ZX \} & 15 = \{ XX, YZ, ZY \} . \end{array} 
\end{equation}

\noindent After careful inspection one finds that the unitary operator basis can 
be divided into disjoint petals in 6 distinct ways, namely
\begin{equation} \begin{array}{lll} \{ 1,2,11,13,14 \}  
& \{ 4,6,8,10,15 \}  & \{ 2,3,7,8,12 \} \\
\{ 1, 3,9,10,15\}  &  
\{ 4,5,11,12,14 \} & \{ 5, 6, 7, 9, 13 \} \ . \end{array}      
\end{equation}

\begin{figure}[h]
        \centerline{ \hbox{
                 \epsfig{figure=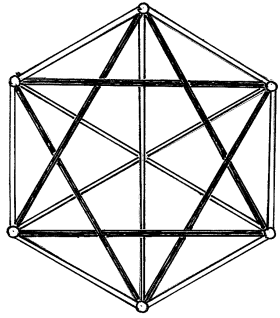, width=38mm}}}
        \caption{The 15 edges of the complete 6-graph represent the bases 
defined by the altogether 15 maximal abelian subgroups of $H(2)\times H(2)$. 
The vertices are complete sets of MUB. The 5 edges meeting at a vertex 
represent the bases in that set. The bases occurring in the 
Mermin square are painted black.}
        \label{fig:David}
\end{figure}

\noindent So we have six flowers, each of which contains exactly two Mermin petals 
(not by accident, as Problem 12.7 
reveals). The pattern 
is summarized in Figure \ref{fig:David}.

This construction can be generalized to any prime power dimension $N = p^K$, The 
multipartite Heisenberg group $H(p)^{\otimes K}$ gives many interlocking flowers,  
and hence many complete sets of MUB. The finite Galois fields are useful here: the 
formulas can be made to look quite similar to (\ref{ivanovic}), except that the 
field theoretic trace is used in the exponents of $\omega$. We do not go into details here, but mention only that
Complete sets of MUB were constructed for prime power dimensions by Wootters 
and Fields (1989) \cite{WF89}. Calderbank et al. \cite{CCKS97} gave a more complete list, still relying on Heisenberg groups. 
All known constructions are unitarily equivalent 
to one on their list \cite{GoRo09}.

The question whether all complete sets of MUB 
can be transformed into each other by means of some unitary transformation arises 
at this point. It is a difficult one. For $N \leq 5$ one can show that all complete 
sets are unitarily equivalent to each other, but for $N = 2^5$ this is no longer 
true. In this case the 5--partite Heisenberg group can be partitioned into flowers 
in several unitarily inequivalent ways.
This was noted by Kantor, who has also reviewed the full story \cite{Kan12}.

What about dimensions that are not prime powers? Unitary operator bases exist in 
abundance, and defy classification, so it is not easy to judge whether there may 
be a flower among them. Nice error bases are easier to deal with (because groups 
can classified). There are Heisenberg groups in every dimension, and if the dimension is 
$N = p_1^{K_1}p_2^{K_2} \dots p_n^{K_n}$ one can use them to construct 
${\rm min}_i(p_i^{K_i})+1$ MUB in the composite dimension $N$ 
\cite{KlaRoe04, Zau99}. 
What is more, it is known that this is the 
largest number one can obtain from the partitioning of any nice error basis into 
petals \cite{Asc07}. However, the story does not end there, because---making use 
of the combinatorial concepts introduced in the next section---one can show that 
in certain square dimensions larger, but still incomplete, sets do exist. 
In particular, in dimension $N = 26^2 = 2^213^2$ group theory would suggest at most 
5 MUB, but Wocjan and Beth found a set of 6 MUB. Moreover, in square dimensions 
$N = m^2$ the number of MUB grows at least as fast as $m^{1/14.8}$ (with finitely 
many exceptions) \cite{WoBe05}. 

This leaves us somewhat at sea. We do not know whether a complete set of MUB exists 
if the dimension is $N = 6, 10, 12, \dots$. How is the question to be settled? 
Numerical computer searches with carefully defined error bounds could settle the 
matter in low dimensions, but the only case that has been investigated 
in any depth is that of $N = 6$. In this case there is convincing evidence 
that at most 3 MUB can be found, but still no 
proof. (Especially convincing evidence was found by Brierley and Weigert 
\cite{BrW08}, Jaming et al. \cite{JMMSW09}, and Raynal et al. \cite{RLE11}. See 
also Grassl \cite{Gr04}, who studies the set of vectors unbiased relative to 
both the computational and Fourier basis. There are 48 such vectors).
And $N = 6$ may be a very special case. 

A close relative of the MUB existence problem arises in Lie algebra theory, and 
is unsolved there as well.
But at least it received 
a nice name: the Winnie--the--Pooh problem. The reason cannot be fully rendered 
into English \cite{KT94}.

 One can imagine that it has do with harmonic (Fourier) 
analysis 
\cite{Ma12}. Or perhaps with symplectic topology: there exists an elegant 
geometrical theorem which says that given any two bases in ${\mathbb C}^N$, 
not necessarily unbiased, there always exist at least $2^{N-1}$ vectors that are 
unbiased relative to both bases. But there is no information about whether 
these vectors form orthogonal bases (and in non--generic cases the vectors may 
coincide). (See \cite{AB15} for a description of this theorem, and 
\cite{Ar00} for a description of this area of mathematics).
 We offer these 
suggestions as hints, and return to a brief summary of known facts at the 
end of Section \ref{sec:finite}. Then we will have introduced some 
combinatorial ideas which---whether they 
have any connection to the MUB existence problem or not---have a number of 
applications to physics.

\section{Finite geometries and discrete Wigner functions} 
\label{sec:finite}

A combinatorial structure underlying the usefulness of mutually unbiased 
bases is that of {\it finite affine planes}. A finite plane is just like 
an ordinary plane, but the number of its points is finite. A finite affine 
plane contains lines too, which by definition are subsets of the set of 
points. It is required that for any pair of points there is a unique line 
containing them. It is also required that for every point not contained 
in a given line there exists a unique line having no point in common with 
the given line. (Devoted students of Euclid will recognize this as the 
Parallel Axiom, and will also recognize that disjoint lines deserve to 
be called parallel.) Finally, to avoid degenerate cases, it is required 
that there are at least two points in each line, and that there are at 
least two distinct lines. With these three axioms one can prove that two 
lines intersect either exactly once or not at all, and also that for every 
finite affine plane there exists an integer $N$ such that

\smallskip

(i) there are $N^2$ points, 

(ii) each line contains $N$ points,

(iii) each point is contained in $N+1$ lines, 

(iv) and there are altogether $N+1$ sets of $N$ disjoint lines.

\smallskip

\noindent  The proofs of these theorems are exercises in pure combinatorics \cite{Ben95}, 
and appear at first glance quite unconnected to the geometry of quantum states. 

It is much harder to decide whether a finite affine plane of order $N$ actually 
exists. If $N$ is a power of a prime number finite planes can be constructed 
using coordinates, just like the ordinary plane (where lines are defined by 
linear equations in the coordinates that label the points), with the difference 
that the coordinates are elements of a finite field of order $N$. Thus a point 
is given by a pair $(x,y)$, where $x$, $y$ belong to the finite field, and a 
line consists of all points obeying either $y = ax+b$ or $x = c$, where $a,b,c$ 
belong to the field. 
This is not quite the end of the story of the finite affine planes because examples 
have been constructed that do not rely on finite fields, but the order $N$ 
of all these examples is a power of some prime number. Whether there exist finite 
affine planes for any other $N$ is not known. 

Let us go a little deeper into the combinatorics, before we explain what it has 
to do with us. A finite plane can clearly be thought of as a grid of $N^2$ points, 
and its rows and columns provide us with two sets of $N$ disjoint or parallel lines, 
such that each line in one of the sets intersect each line in the other set exactly 
once. But what of the next set of $N$ parallel lines? We can label its lines with 
letters from an alphabet containing $N$ symbols, and the requirement that any two 
lines meet at most once translates into the observation that finding the third set 
is equivalent to finding a Latin square. As we saw in Section \ref{sec:Werner}, 
there are many Latin squares to choose from. The difficulty comes in the next 
step, when we ask for two Latin squares describing two different sets of parallel 
lines. Use Latin letters as the alphabet for the first, and Greek letters for 
the second. Then each point in the array will be supplied with a pair of letters, 
one Latin and one Greek. Since two lines are forbidden to meet more than once 
a given pair of letters, such as $(A, \alpha)$ or $(B, \gamma )$, 
is allowed to occur only once in the array. In other words the letters from the 
two alphabets serve as alternative coordinates for the array. Pairs of Latin 
squares enjoying this property are known as {\it Graeco-Latin} or {\it orthogonal 
Latin squares}. For $N = 3$ it is easy to find Graeco-Latin pairs, such as  
\begin{equation} \left( \ \ \begin{array}{|c|c|c|} \hline A & B & C \\ 
\hline B & C & A \\ \hline 
C & A & B \\ \hline \end{array} \ , \ \begin{array}{|c|c|c|} \hline \alpha & \beta & 
\gamma \\ \hline \gamma & \alpha & \beta \\ \hline 
\beta & \gamma & \alpha \\ \hline \end{array} \ \ \right)
\ = \ 
\begin{array}{|c|c|c|} \hline 
{ A\alpha}  & B\beta& {C\gamma}\\ \hline
{ B\gamma} & C\alpha & { A\beta}\\ \hline
{C\beta}    & A\gamma  &{B\alpha} \\ \hline
\end{array} \ . \end{equation}

\noindent An example for $N = 4$, using alphabets that may appeal to bridge players, 
is

\begin{equation} {\small \begin{array}{|cccc|} \hline 
{ A\spadesuit}  & K\clubsuit& {Q\diamondsuit} & J\heartsuit \\ 
{ K\heartsuit} & A\diamondsuit & { J\clubsuit}& Q\spadesuit\\ 
{Q\clubsuit}    & J\spadesuit  &{A\heartsuit} & K\diamondsuit \\ 
J\diamondsuit & Q\heartsuit & K\spadesuit & A\clubsuit \\ \hline
\end{array} } \ . \end{equation}

\noindent Graeco-Latin pairs can be found for all choices of $N>2$ except 
(famously) $N = 6$. 
This is so even though Table \ref{tab:latin} shows that there 
is a very large supply of Latin squares for $N = 6$. 
The story behind these non-existence results goes back to Euler, who was concerned with arranging 36 officers
 belonging to 6 regiments, 6 ranks, and 6 arms, in a square.

To define a complete affine plane, with $N+1$ sets of parallel 
lines, requires us to find a set of $N-1$ {\it mutually orthogonal Latin 
squares}, or {\it MOLS}.  
For $N = 6$ and $N = 10$ this is impossible, and in fact an 
infinite number of possibilities (beginning with $N = 6$) are ruled out by the 
{\it Bruck-Ryser theorem}, which says that if an affine plane of order $N$ exists, 
and if $N = 1$ or 2 modulo 4, then $N$ must be the sum of two squares. Note that 
10 is a sum of two squares, but this case has been ruled out by different 
means. Lam describes the computer based non-existence proof for $N = 10$ in a 
thought-provoking way \cite{Lam91}.

 If $N = p^k$ 
for some prime number $p$ a solution can easily be found using analytic geometry 
over finite fields. There remain an infinite number of instances, beginning 
with $N = 12$, for which the existence of a finite affine plane is an open 
question.
See the books by Bennett \cite{Ben95}, and Stinson \cite{Sti04}, for more 
information, and for proofs of the statements we have made so far.

At this point we recall that complete sets of MUB exist when $N = p^k$, but 
quite possibly not otherwise. Moreover such a complete set is naturally described 
by $N+1$ sets of $N$ vectors. The total number of vectors is the same as the 
number of lines in a finite affine plane, so the question is if we can somehow 
associate $N^2$ `points' to a complete set of MUB, in such a way that the 
incidence structure of the finite affine plane becomes useful. One way to do this 
is to start with the picture of a complete set of MUB as a polytope in Bloch space. 
We do not have to assume that a complete set of MUB exists. We simply introduce 
$N(N+1)$ Hermitian matrices $P_\nu$ of unit trace, 
\index{polytope!complementarity} 
denoted $P_v$, and obeying
\begin{equation} \mbox{Tr}P_vP_{v'} = \left\{ \begin{array}{lll} 1 & \mbox{if} & v = v' \\ 
1/N & \mbox{if} & v \ \mbox{and} \ v' \ \mbox{sit in different planes} \\ 
0 & \mbox{if} & v \neq v' \ \mbox{sit in the same plane} \end{array} \right. 
\label{vertices} \end{equation}

\noindent The condition that Tr$P_v^2 = 1$ ensures that $P_v$ lies on the outsphere of the 
set of quantum states. If their eigenvalues are non-negative these are projectors, and 
then they actually are quantum states, but this is not needed for the definition of the 
polytope. To understand the face structure of the polytope 
we begin by noting that the 
convex hull of one vertex from each of the $N+1$ individual simplices forms a face.  
(This is fairly obvious, and anyway we are just about to prove it.) Using the 
matrix representation of the vertices we can then form the Hermitian unit trace matrix 
\begin{equation} A_f = \sum_{\rm face} P_v - {\mathbbm 1} \ , \label{Af} \end{equation}

\noindent where the sum runs over the $N+1$ vertices in the face. This is called a 
{\it face point operator} (later to be subtly renamed as a {\it phase point operator}). 
If $N = 3$ we can think of it pictorially as 
$\triangle \hspace{-3.8mm} 
{\ }^\bullet \hspace{2.3mm} \triangle \hspace{-3.8mm} {\ }^\bullet \hspace{2.3mm}  
\triangle \hspace{-3.8mm} {\ }^\bullet \hspace{2.3mm} 
\triangle \hspace{-3.8mm} {\ }^\bullet \hspace{2.5mm}$, say---each triangle represents a basis. 
See Figure \ref{fig:mub1}. It is easy to see that $0 \leq \mbox{Tr}\rho A_f\leq 1$ 
for any matrix $\rho$ that lies in the complementarity polytope, which means that the 
latter is confined between two parallel hyperplanes. There is 
a facet defined by $\mbox{Tr}\rho A_f = 0$ (pictorially, this would be 
$\triangle \hspace{-5mm} {\ }_{\bullet} \hspace{-0.8mm} 
{\ }_{\bullet} \hspace{2.3mm} \triangle \hspace{-5mm} {\ }_{\bullet} 
\hspace{-0.8mm} {\ }_{\bullet} \hspace{2.3mm} \triangle \hspace{-5mm} 
{\ }_{\bullet} \hspace{-0.8mm} {\ }_{\bullet} \hspace{2.3mm} 
\triangle \hspace{-5mm} {\ }_{\bullet} \hspace{-0.8mm} {\ }_{\bullet} 
\hspace{1mm}$) 
and an opposing face containing one 
vertex from each simplex. Every vertex is included in one of these two faces. 
There are $N^{N+1}$ operators $A_f$ altogether, and equally many facets.   

The idea is to select $N^2$ phase point 
operators and use them to represent the points of an affine plane. The $N+1$ 
vertices $P_\nu$ that appear in the sum (\ref{Af}) are to be regarded as the $N+1$ 
lines passing through the point $A_f$. A set of $N$ parallel lines in the affine plane 
will represent a complete set of orthonormal projectors.  

\begin{figure}[ht]
        \centerline{ \hbox{
                 \epsfig{figure=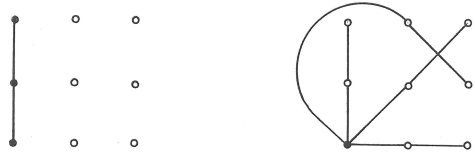, width=65mm}}}
        \caption{All the points on a line determine a $P_v$, and all the lines 
through a point determine an $A_{ij}$. In fact we are illustrating eqs. 
(\ref{plansummor}).}
        \label{fig:mub4}
\end{figure}

To do so, recall that each $A_f$ is defined by picking one $P_v$ from each basis. 
Let us begin by making all possible choices from the first two, and arrange 
them in an array:
\begin{equation} \begin{array}{ccccc} \triangle \hspace{-3.8mm} {\ }^\bullet 
\hspace{3.2mm} \triangle \hspace{-2.6mm} {\ }_{\bullet} \hspace{2mm} 
\triangle \ \triangle & \ 
& \triangle \hspace{-5mm} {\ }_{\bullet} \hspace{4mm} 
\triangle \hspace{-2.6mm} {\ }_{\bullet} \hspace{2mm} \triangle \ \triangle & 
\ & \triangle \hspace{-2.6mm} {\ }_{\bullet} \hspace{2mm} 
\triangle \hspace{-2.6mm} {\ }_{\bullet} \hspace{2mm} \triangle \ \triangle \\
\ & \ & \ & \ & \ \\
\triangle \hspace{-3.8mm} {\ }^\bullet 
\hspace{3.2mm} \triangle \hspace{-5mm} {\ }_{\bullet} \hspace{4.2mm} 
\triangle \ \triangle 
& \ & \triangle \hspace{-5mm} {\ }_{\bullet} 
\hspace{4.2mm} \triangle \hspace{-5mm} {\ }_{\bullet} \hspace{4.2mm} 
\triangle \ \triangle & \ &  
\triangle \hspace{-2.6mm} {\ }_{\bullet} \hspace{2mm} 
\triangle \hspace{-5mm} {\ }_{\bullet} \hspace{4.2mm} 
\triangle \ \triangle \\ 
\ & \ & \ & \ & \ \\
\triangle \hspace{-3.8mm} {\ }^\bullet 
\hspace{3.2mm} \triangle \hspace{-4mm} {\ }^\bullet 
\hspace{3.2mm} \triangle \ \triangle 
& \ & \triangle \hspace{-5mm} {\ }_{\bullet} \hspace{4mm} 
\triangle \hspace{-4mm} {\ }^\bullet 
\hspace{3.2mm} \triangle \ \triangle & \ & 
\triangle \hspace{-2.6mm} {\ }_{\bullet} \hspace{2mm} 
\triangle \hspace{-4mm} {\ }^\bullet 
\hspace{3.2mm} \triangle \ \triangle \end{array} 
\end{equation}

\noindent We set $N = 3$---enough to make the idea come through---in this illustration. 
Thus the $N+1$ simplices in the totally orthogonal $(N-1)$-planes appear as 
four triangles, two of which have been used up to make the array. We use the vertices 
of the remaining $N-1$ simplices to label the lines in the remaining $N-1$ 
pencils of parallel lines. To ensure that non-parallel lines intersect exactly 
once a pair such as $\triangle \hspace{-2.6mm} {\ }_{\bullet} \hspace{2mm} 
\triangle \hspace{-5mm} {\ }_{\bullet} \ $ (picked from any two out of the four 
triangles) must occur exactly once in the array. This problem can be solved 
because  an affine plane of order $N$ is presumed to exist. One solution is  
\begin{equation} \begin{array}{ccccc} \triangle \hspace{-3.8mm} {\ }^\bullet 
\hspace{3.2mm} \triangle \hspace{-2.6mm} {\ }_{\bullet} \hspace{2mm} 
\triangle \hspace{-5mm} {\ }_{\bullet} \hspace{4mm} 
\triangle \hspace{-2.6mm} {\ }_{\bullet} \hspace{2mm} & \ 
& \triangle \hspace{-5mm} {\ }_{\bullet} \hspace{4mm} 
\triangle \hspace{-2.6mm} {\ }_{\bullet} \hspace{2mm} 
\triangle \hspace{-2.6mm} {\ }_{\bullet} \hspace{2mm} 
\triangle \hspace{-4mm} {\ }^\bullet \hspace{3.2mm} & 
\ & \triangle \hspace{-2.6mm} {\ }_{\bullet} \hspace{2mm} 
\triangle \hspace{-2.6mm} {\ }_{\bullet} \hspace{2mm} 
\triangle \hspace{-4mm} {\ }^\bullet \hspace{3.2mm} 
\triangle \hspace{-5mm} {\ }_{\bullet} \hspace{4.2mm} \\
\ & \ & \ & \ & \ \\
\triangle \hspace{-3.8mm} {\ }^\bullet 
\hspace{3.2mm} \triangle \hspace{-5mm} {\ }_{\bullet} \hspace{4.2mm} 
\triangle \hspace{-2.6mm} {\ }_{\bullet} \hspace{2mm} 
\triangle \hspace{-5mm} {\ }_{\bullet} \hspace{4.2mm} 
& \ & \triangle \hspace{-5mm} {\ }_{\bullet} 
\hspace{4.2mm} \triangle \hspace{-5mm} {\ }_{\bullet} \hspace{4.2mm} 
\triangle \hspace{-4mm} {\ }^\bullet 
\hspace{3.2mm} 
\triangle \hspace{-2.6mm} {\ }_{\bullet} \hspace{2mm} & \ &  
\triangle \hspace{-2.6mm} {\ }_{\bullet} \hspace{2mm} 
\triangle \hspace{-5mm} {\ }_{\bullet} \hspace{4.2mm} 
\triangle \hspace{-5mm} {\ }_{\bullet} \hspace{4mm} 
\triangle \hspace{-4mm} {\ }^\bullet \hspace{3.2mm} \\ 
\ & \ & \ & \ & \ \\
\triangle \hspace{-3.8mm} {\ }^\bullet 
\hspace{3.2mm} \triangle \hspace{-4mm} {\ }^\bullet 
\hspace{3.2mm} \triangle \hspace{-4mm} {\ }^\bullet \hspace{3.2mm} 
\triangle \hspace{-4mm} {\ }^\bullet \hspace{3.2mm} 
& \ & \triangle \hspace{-5mm} {\ }_{\bullet} \hspace{4mm} 
\triangle \hspace{-4mm} {\ }^\bullet \hspace{3.2mm} 
\triangle \hspace{-5mm} {\ }_{\bullet} \hspace{4mm} 
\triangle \hspace{-5mm} {\ }_{\bullet} \hspace{4mm} & \ & \ \ 
\triangle \hspace{-2.6mm} {\ }_{\bullet} \hspace{2mm} 
\triangle \hspace{-4mm} {\ }^\bullet \hspace{3.2mm} 
\triangle \hspace{-2.6mm} {\ }_{\bullet} \hspace{2mm}  
\triangle \hspace{-2.6mm} {\ }_{\bullet} \hspace{2mm} 
\hspace{3.2mm} \end{array} 
\label{afplane}
\end{equation}

\noindent We have now singled out $N^2$ face point operators for attention, 
and the combinatorics of the affine plane guarantees that any pair of them 
have exactly one $P_v$ in common. Equation (\ref{vertices}) then enables us 
to compute that 
\begin{equation} \mbox{Tr}A_fA_{f'} = N\delta_{f,f'} \ . \end{equation}

\noindent This is a regular simplex in dimension $N^2-1$. 

We used only $N^2$ out of the $N^{N+1}$ face point operators for this construction, 
but a little reflection shows that the set of all of them can be divided into 
$N^{N-1}$ disjoint face point operator simplices. In effect we have inscribed 
the complementarity polytope into this many regular simplices, which is an 
\index{polytope!complementarity}
interesting datum about the former. Each such simplex forms an orthogonal operator 
basis, although not necessarily a unitary operator basis. Let us focus on one of them, 
and label the operators occurring in it as $A_{i,j}$. It is easy to see that eq. 
(\ref{Af}) can be rewritten (in the language of the affine plane), and supplemented, 
so that we have the two equations 
\begin{equation} P_v = \frac{1}{N}\sum_{{\rm line}\ v} A_{i,j} \ , \hspace{10mm} 
A_{i,j} = \sum_{{\rm point}\ (i,j)}P_v - {\mathbbm 1} \ . \label{plansummor} \end{equation}

\noindent The summations extend through all points on the line, respectively all lines passing 
through the point. Using the fact that the phase point operators form an operator 
basis we can then define a discrete Wigner function by 
\begin{equation} W_{i,j} = \frac{1}{N}\mbox{Tr}A_{i,j}\rho \ . \end{equation}

\noindent Knowledge of the $N^2$ real numbers $W_{i,j}$ is equivalent to knowledge of the 
density matrix $\rho$. Each line in the affine plane is now associated with the number 
\begin{equation} p_v = \sum_{{\rm line}\ v}W_{i,j} = \mbox{Tr}P_v\rho \ . \end{equation}

\noindent Clearly the sum of these numbers over a pencil of parallel lines equals unity. 
However, so far we have used only the combinatorics of the complementarity polytope, and 
we have no right to expect that the operators $P_v$ have positive spectra. They will be 
projectors onto pure states if and only if the complementarity polytope has been 
inscribed into the set of density matrices, which is a difficult thing to achieve. If it 
is achieved we conclude that $p_v \geq 0$, and then we have an 
elegant discrete Wigner function---giving all the correct marginals---on 
our hands \cite{Wo87}. It will receive further polish 
in the next section. 

Meanwhile, now that we have the concepts of mutually orthogonal Latin squares 
and finite planes on the table, we 
can discuss some interesting but rather abstract analogies to the MUB existence 
problem. Fix $N=p_1^{K_1}p_2^{K_2}\dots p_n^{K_n}$. Let $\#_{\rm MUB}$ be the 
number of MUB, and let $\#_{\rm MOLS}$ be the number of MOLS.
(We need only $N-1$ MOLS to construct a finite affine plane.) Then 
\begin{equation} \begin{array}{l} {\rm min}_i(p_i^{K_i}) -1 \leq \#_{\rm MOLS} 
\leq N-1 \\ \\ {\rm min}_i(p_i^{K_i}) + 1 \leq \#_{\rm MUB} \leq N+1 \ . 
\end{array} \end{equation}

\noindent The lower bound for MOLS is known as the MacNeish bound \cite{Ma21}. 
Moreover, if $N = 6$, we know that Latin squares cannot occur in orthogonal pairs, 
and we believe that there exist only three MUB. Finally, it is known that if there exist 
$N-2$ MOLS there necessarily exist $N-1$ MOLS, and if there exist $N$ MUB there 
necessarily exist $N+1$ MUB \cite{We13}. This certainly encourages the speculation that 
the existence problem for finite affine planes is related to the existence 
problem for complete sets of MUB in some unknown way. However, the idea fades 
a little if one looks carefully into the details \cite{PPGB10, WD10}.

\section{Clifford groups and stabilizer states}
\label{sec:Clifford}

To go deeper into the subject we need to introduce the {\it Clifford group}. 
\index{group!Weyl--Heisenberg}
\index{group!Clifford}
We use the displacement operators $D_{\bf p}$ from Section \ref{sec:uob} to 
describe the Weyl--Heisenberg group. By definition the Clifford group consists 
of all unitary operators $U$ such that 
\begin{equation} UD_{\bf p}U^{-1} \sim D_{f({\bf p})} \ , \end{equation}

\noindent where $\sim$ means `equal up to phase factors'. (Phase factors will 
appear if $U$ itself is a displacement operator, which is allowed.) Thus we ask 
for unitaries that permute the displacement operators, so that the conjugate of 
an element of the Weyl-Heisenberg group is again a member of the Weyl--Heisenberg 
group. The technical term for this is that the Clifford group is the 
{\it normalizer} of the Weyl-Heisenberg group within 
the unitary group, or that the Weyl--Heisenberg group is an {\it invariant subgroup}
of the Clifford group. If we change $H(N)$ into a non-isomorphic multipartite 
Heisenberg group we obtain another Clifford group, but at first we stick with 
$H(N)$.
(The origin of the name `Clifford group' is a little unclear to us. It 
seems to be connected to Clifford's gamma matrices rather than to Clifford 
himself \cite{BRW60}).

The hard part of the argument is to show that 
\begin{equation} U_GD_{\bf p}U^{-1}_G \sim D_{G{\bf p}} \ , \label{slsl} \end{equation}

\noindent where $G$ is a two-by-two matrix with entries in ${\mathbbm Z}_{\bar{N}}$, the point 
being that the map ${\bf p} \rightarrow f({\bf p})$ has to be linear \cite{App05}. We take 
this on trust here. To see the consequences we 
return to the group law (\ref{Hgrouplaw}). In the exponent of 
the phase factor we encounter the symplectic form 
\begin{equation} \Omega ({\bf p},{\bf q}) = p_2q_1 - p_1q_2 \ . \end{equation}

\noindent When strict equality holds in eq. (\ref{slsl}) it follows from the group law that  
\begin{equation} U_GD_{\bf p}D_{\bf q}U^{-1}_G = \tau^{\Omega ({\bf p},{\bf q})}
UD_{{\bf p} + {\bf q}}U^{-1}_G \hspace{2mm} \Rightarrow \hspace{2mm} 
D_{G{\bf p}}D_{G{\bf q}} = \tau^{\Omega ({\bf p},{\bf q})}D_{G({\bf p}+{\bf q})} 
\ .  \end{equation}

\noindent On the other hand we know that 
\begin{equation}
D_{G{\bf p}}D_{G{\bf q}} = \tau^{\Omega (G{\bf p},G{\bf q})}D_{G({\bf p}+{\bf q})} 
\ .  \end{equation}

\noindent Consistency requires that 
\begin{equation} \Omega ({\bf p},{\bf q}) = \Omega (G{\bf p},G{\bf q}) \hspace{4mm}  
\mbox{mod} \ \bar{N} \ . \label{marcus1} \end{equation}

\noindent The two by two matrix $G$ must leave the symplectic form invariant. 
The arithmetic in the exponent is modulo $\bar{N}$, where $\bar{N} = N$ if $N$ is odd and 
$\bar{N} = 2N$ if $N$ is even. We deplored this unfortunate complication in even 
dimensions already in Section \ref{sec:uob}.

Let us work with explicit matrices 
\begin{equation} \Omega = \left( \begin{array}{cc} 0 & - 1 \\ 1 & 0 \end{array} 
\right) \ , \hspace{8mm} 
G = \left( \begin{array}{cc} \alpha & \beta \\ \gamma & \delta \end{array}\right) 
\ , \hspace{8mm} \alpha , \beta , \gamma , \delta \in {\mathbbm Z}_{\bar{N}} \ . 
\label{SL2} \end{equation}

\noindent Then eq. (\ref{marcus1}) says that 
\begin{equation} \left( \begin{array}{cc} 0 & -1 \\ 1 & 0 \end{array}\right) = 
\left( \begin{array}{cc} 
0 & -\alpha \delta + \beta \gamma \\ \alpha \delta - \beta \gamma & 0 
\end{array}\right) \ . \end{equation}

\noindent Hence the matrix $G$ must have determinant equal to 1 modulo $\bar{N}$. 
Such matrices form the group $SL(2, {\mathbbm Z}_{\bar{N}})$, where ${\mathbbm Z}_{\bar{N}}$ 
stands for the ring of integers modulo $\bar{N}$. It is also known as a {\it symplectic 
group}, because it leaves the symplectic form invariant. 
\index{group!symplectic}

The full structure of the Clifford group $C(N)$ is complicated 
by the phase factors, and rather difficult to describe in general. 
Things are much simpler when $\bar{N} = N$ is odd, so let us restrict 
our description to this case.
(A complete, and clear, account 
of the general case is given by Appleby \cite{App05}).
 Then the symplectic 
group is a subgroup of the Clifford group. Another subgroup is evidently 
the Weyl--Heisenberg group itself. Moreover, if we consider the Clifford 
group modulo its centre, $C(N)/I(N)$, that is to say if we identify group 
elements differing only by phase factors---which we would naturally do if 
we are interested only in how it transforms the quantum states---then we 
find that $C(N)/I(N)$ is a semi-direct product of the symplectic rotations 
given by $SL(2, {\mathbbm Z}_{N})$ and the translations given by $H(N)$ 
modulo its centre. 

In every case---also when $N$ is even---the unitary representation of the Clifford group is 
uniquely determined by the unitary representation of the Weyl-Heisenberg group.
The easiest case to describe is that when $N$ is an odd prime number $p$. Then the 
symplectic group is defined over the finite field ${\mathbbm Z}_p$ consisting of integers 
modulo $p$, and it contains $p(p^2-1)$ elements altogether. Insisting that there exists 
a unitary matrix $U_G$ such that $U_G D_{\bf p} U_G^{-1} = D_{G{\bf p}}$ we are led to the 
representation 
\begin{equation}  G = \left( \begin{array}{cc} \alpha & \beta \\ \gamma 
& \delta \end{array}\right) \hspace{1mm} \rightarrow \hspace{1mm} 
\left\{ \begin{array}{ll} 
U_G = \frac{e^{i\theta}}{\sqrt{p}}\sum_{i,j}\omega^{\frac{1}{2\beta}
(\delta i^2 - 2ij + \alpha j^2)}|i\rangle \langle j| & \ \beta \neq 0 \\ \\ 
U_G = \pm \sum_j\omega^{\frac{\alpha \gamma}{2}j^2}|\alpha j\rangle \langle j| 
& \ \beta = 0 \ . \end{array} \right. \label{marcusa} \end{equation}    

\noindent In these formulas `$1/\beta$' stands for the multiplicative inverse 
of the integer $\beta$ in arithmetic modulo $p$ (and since $1/2$ occurs it is 
obvious that special measures must be taken if $p=2$). An 
overall phase factor is left undetermined: it can be pinned down by insisting on 
a faithful representation of the group \cite{Neu02, App09}, but in many situations 
it is not needed. It is noteworthy that the representation matrices are either 
complex Hadamard matrices, or monomial matrices. 
It is interesting to see how they act on the 
set of mutually unbiased bases. In the affine plane a symplectic transformation 
takes lines to lines, and indeed parallel lines to parallel lines. If one works 
this out one finds that the symplectic group acts like M\"obius transformations 
\index{transformation!M\"obius}
on a projective line whose points are the individual bases. See Problem 12.5.

Even though $N = 2$ is even, it is the easiest case to understand. The collineation group $C(2)/I(2)$ 
is just the group of rotations that transforms the polytope formed by the MUB states into itself, 
or in other words it is the symmetry group $S_4$ of the octahedron. In higher prime dimensions the 
Clifford group yields only a small subgroup of the symmetry group of the complementarity polytope. 
When $N$ is a composite number, and especially if $N$ is an even composite 
number, there are some significant complications for which we refer elsewhere \cite{App05}. 
These complications do sit in the background in Section \ref{sec:SICs}, where 
the relevant group is the {\it extended Clifford group}
obtained by allowing also two-by-two matrices of determinant $\pm 1$. In Hilbert space this 
doubling of the group is achieved by representing the extra group elements by anti-unitary transformations \cite{App05}. 

To cover MUB in prime power dimensions we need to generalize in a different 
direction. The relevant Heisenberg group is the multipartite Heisenberg group. 
We can still define the Clifford group as 
the normalizer of this Heisenberg group. We recall that the latter contains many 
maximal abelian subgroups, and we refer to the joint eigenvectors of these 
subgroups as {\it stabilizer states}. The Clifford group acts by permuting the 
stabilizer states, and every such permutation can be built as a sequence of 
operations on no more than two qubits (or quNits as the case may be) at a time. 
In one standard blueprint for universal quantum computing \cite{Go97}, the quantum 
computer is able to perform such permutations in a fault--tolerant way, and the 
stabilizer states play a role reminiscent of that played by the separable 
states (to be defined in Chapter 16) 
in quantum communication.

The total number $M$ of stabilizer states in $N = p^K$ dimensions is 
\begin{equation} M = p^K\prod_{i=1}^K(p^i+1) \ . \end{equation}

\noindent Dividing out a factor $p^K$ we obtain the number of maximal abelian subgroups 
of the Heisenberg group. In dimension $N = 2^2$ there are altogether 60 stabilizer 
states forming 15 bases and 6 interlocking complete sets of MUB, 
because there are 6 different ways in which the group $H(2)\times H(2)$ 
can be displayed as a flower. See Figure \ref{fig:David}. 

The story in higher dimensions 
is complicated by the appearance of complete sets that fail to be unitarily equivalent 
to each other. We must refer elsewhere for the details \cite{Kan12}, but it is 
worth remarking that, for the `canonical' choice of a complete set written down by 
Ivanovi\'c \cite{Iv81} and by Wootters and Fields \cite{WF89}, there exists 
a very interesting subgroup of the Clifford group leaving this set invariant. 
It is known as the {\it restricted Clifford group} \cite{App09}, and has an elegant 
description in terms of finite fields. 
\index{group!restricted Clifford}
Moreover (with an exception in dimension 3) the set of vectors that make up this set of 
MUB is distinguished by the property that it provides the smallest orbit under 
this group \cite{Zh15}. 
For both Clifford groups, the quotient of their collineation groups with the discrete 
translation group provided by their Heisenberg groups is a symplectic group. If we start 
with the full Clifford group the symplectic group acts on a 
$2K$-dimensional vector space over ${\mathbbm Z}_p$, while in the case of the  
restricted Clifford group it can be identified with the group 
$SL(2,{\mathbbm F}_{p^K})$ acting on a 2-dimensional vector space over the 
finite field ${\mathbbm F}_{p^K} = GF(p^K)$ \cite{Gro06}.  

Armed with these group theoretical facts we can return to the subject of discrete Wigner 
functions. If we are in a prime power dimension it is evident that we can produce 
a phase point operator simplex by choosing any phase point operator $A_f$, and act 
on it with the appropriate Heisenberg group. But we can ask for more. We can ask 
for a phase point operator simplex that transforms into itself when acted on by 
the Clifford group. If we succeed, we will have an affine plane that behaves like 
a true phase space, since it will be equipped with a symplectic structure. This turns 
out to be possible in odd prime power dimensions, but not quite possible when the 
dimension is even. We confine ourselves to odd prime dimensions here. Then the Clifford 
group contains a unique element of order two, whose unitary representative 
we call $A_{0,0}$ \cite{Wo87}. Using eq. (\ref{marcusa}) it is   
\begin{equation} G = \left( \begin{array}{rr} -1 & 0 \\ 0 & -1 \end{array} \right) 
\hspace{5mm} \Rightarrow \hspace{5mm} 
A_{0,0} = U_G = \sum_{i=0}^{N-1} |N-i\rangle \langle i| \ . \end{equation}

\noindent By the way we observe that $A_{0,0}^2 = F$, where $F$ is the Fourier matrix. 
Making use of eqs. (\ref{WHUOB1}-\ref{WHUOB2}) we find  
\begin{equation} A_{0,0} = \sum_{r,s}D_{r,s} \ . \end{equation}

\noindent To perform this sum, split it into a sum over the $N+1$ maximal abelian 
subgroups and subtract ${\mathbbm 1}$ to avoid overcounting. Diagonalize 
each individual generator of these subgroups, say
\begin{equation} D_{0,1} = Z = \omega^a\sum_a |0,a\rangle \langle 0,a| \hspace{5mm} \Rightarrow 
\hspace{5mm} \sum_{i=0}^{N-1}Z^i = N|0,0\rangle \langle 0,0| \ . \end{equation}

\noindent All subgroups work the same way, so this is enough. We conclude that 
\begin{equation} A_{0,0} = 
\sum_{z}|z,0\rangle \langle z,0| - {\mathbbm 1} \ . \end{equation}

\noindent (The range of the label $z$ is extended to cover also the 
bases that we have labelled by $0$ and $\infty$.) Since we are picking one projector 
from each of the $N+1$ bases this is in fact a phase point operator.

Starting from $A_{0,0}$ we can build a set of $N^2$ order two phase point operators 
\begin{equation} A_{r,s} = D_{r,s}AD_{r,s}^{-1} \ . \end{equation}

\noindent Their eigenvalues are $\pm 1$, so these operators are both Hermitian and 
unitary. The dimension $N$ is odd, so we can write $N = 2m-1$. Each phase 
point operator splits Hilbert space into a direct sum of eigenspaces, 
\begin{equation} {\cal H}_N = {\cal H}_m^{(+)} \oplus {\cal H}_{m-1}^{(-)} \ . 
\end{equation}

\noindent Altogether we have $N^2$ subspaces of dimension $m$, each of 
which contain $N+1$ MUB vectors. Conversely, one can show that each of the $N(N+1)$ 
MUB vectors belongs to $N$ such subspaces. This intersection pattern was said to be 
``{\it une configuration tr\`es-remarquable}'' when it was first discovered
(By Segre (1886) \cite{Se86}, who was studying elliptic normal curves. 
From the present 
point of view it was first discovered by Wootters (1987) \cite{Wo87}).

The operators $A_{r,s}$ form a 
phase point operator simplex which enjoys the twin advantages of being both a unitary 
operator basis and an orbit under the Clifford group. A very satisfactory 
discrete Wigner function can be obtained from it \cite{Wo87, Gro06}. The situation 
in even prime power dimensions is somewhat less satisfactory since covariance 
under the full Clifford group cannot be achieved in this case.

The set of phase point operators forms a particularly interesting unitary operator 
basis, existing in odd prime power dimensions only. Its symmetry group acts on it in 
such a way that any pair of elements can be transformed to any other pair. This is 
at the root of its usefulness: from it we obtain a discrete Wigner function 
on a phase space lacking any kind of scale, just as the ordinary symplectic 
vector spaces used in classical mechanics lack any kind of scale. Moreover (with 
two exceptions, one in dimension two and and one in dimension eight) it is 
uniquely singled out by this property \cite{Zh16}.

\section{Some designs}
\label{sec:designs}

To introduce our next topic let us say something obvious. 
We know that 
\begin{equation} {\mathbbm 1}_N = \frac{1}{N}\sum_{i=1}^{N}|e_i\rangle \langle e_i| 
= \int_{{\mathbbm C}{\bf P}^n} {\rm d}\Omega_\Psi |\Psi \rangle \langle \Psi | \ , 
\end{equation}

\noindent where ${\rm d}\Omega_\Psi$ is the unitarily invariant Fubini--Study measure.  
Let $A$ be any operator acting on ${\mathbbm C}^N$. It follows that 
\begin{equation} \frac{1}{N}\sum_{i=0}^{N-1}\langle e_i|A|e_i\rangle = 
\int_{{\mathbbm C}{\bf P}^n} {\rm d}\Omega_\Psi \langle \Psi|A|\Psi\rangle = 
\langle \langle \Psi|A|\Psi\rangle \rangle_{\rm FS} \ . \label{statmek} \end{equation}

\noindent On the right hand side we are averaging an (admittedly special) 
function over all of ${\mathbbm C}{\bf P}^{N-1}$. On the left hand side we take 
the average of its values at $N$ special points. In statistical 
mechanics this equation allows us to evaluate the average expectation value 
of the energy by means of the convenient fiction that the system is in an energy 
eigenstate---which at first sight is not obvious at all. 

To see how this can be generalized we recall the mean value theorem, which 
says that for every continuous function defined on the closed 
interval $[0,1]$ there exists a point $x$ in the interval such that 
\begin{equation} f(x) = \int_0^1{\rm d}sf(s) \ . \end{equation}

\noindent Although it is not obvious, this can be generalized to the case of 
sets of functions $f_i$ \cite{SeZa84}. Given such a set of functions one can always 
find an {\it averaging set} consisting of $K$ different points $x_I$ such that, for 
all the $f_i$,   
\begin{equation} \frac{1}{K}\sum_{I=1}^Kf_i(x_I) = \int_0^1{\rm d}s f_i(s) \ . 
\end{equation}  

\noindent Of course the averaging set (and the integer $K$) will depend on the 
set of functions $\{ f_i\}$ one wants to average. 
We can generalize even more by replacing the interval with a 
connected space, such as ${\mathbbm C}{\bf P}^{N-1}$, and by replacing the real 
valued functions with, say, the set Hom$(t,t)$ of all complex valued functions 
homogeneous of order $t$ in the homogeneous coordinates and their complex 
conjugates alike. (The restriction on the functions is needed in order to 
ensure that we get functions on ${\mathbbm C}{\bf P}^{N-1}$. Note that the expression 
$\langle \psi|A|\psi \rangle$ belongs to Hom$(1,1)$.) This too can 
always be achieved, with an averaging set being a 
collection of points represented by the unit vectors $|\Psi_I\rangle$, 
$1\leq I \leq K$, for some sufficiently large integer $K$ \cite{SeZa84}. 
We define a {\it complex projective t--design}, or {\it t--design} for short, 
as a collection of unit vectors $\{ |\Psi_I\rangle\}_{I=1}^K$ such that 
\begin{equation} \frac{1}{K}\sum_{I=1}^Kf(|\Psi_I\rangle ) = 
\int_{{\mathbbm C}{\bf P}^n} {\rm d}\Omega_\Psi f(|\Psi\rangle ) \end{equation}

\noindent for all polynomials $f \in \mbox{Hom}(t,t)$ with the components 
of the vector, and their complex conjugates, as arguments. 
Formulas like this are called 
{\it cubature formulas}, since---like quadratures---they give explicit solutions 
of an integral, and they are of practical interest---for many signal processing and 
quantum information tasks---provided that $K$ can be chosen to be reasonably small. 

Eq. (\ref{statmek}) shows that orthonormal bases are 1--designs. More generally, 
every POVM 
\index{POVM}
is a 1--design. Let us also note that functions $f \in$ Hom$(t-1,t-1)$ 
can be regarded as special cases of functions in Hom$(t,t)$, since they can be rewritten 
as $f = \langle \Psi|\Psi\rangle f \in$ Hom$(t,t)$. Hence a $t$--design is automatically 
a $(t-1)$--design. But how do we recognize a $t$--design when we see one?

The answer is quite simple. 
In eq. (7.69) 
we calculated the Fubini--Study average of 
$|\langle \Phi|\Psi\rangle|^{2t}$ for a fixed unit vector $|\Phi \rangle$. 
Now let $\{ |\Psi_I\rangle \}_{I = 1}^K$ be a $t$--design. It follows that 
\begin{equation} 
\frac{1}{K}\sum_J |\langle \Psi_I|\Psi_J \rangle |^{2t} = 
\langle |\langle \Psi_I|\Psi\rangle |^{2t} 
\rangle_{\rm FS} = \frac{t!(N-1)!}{(N-1+t)!} \ . \end{equation} 

\noindent If we multiply by $1/K$ and then sum over $I$ we obtain 
\begin{equation} \frac{1}{K^2}\sum_{I,J}|\langle \Psi_I|\Psi_J\rangle |^{2t} = 
\frac{t!(N-1)!}{(N-1+t)!} \ . \label{designtheorem} \end{equation} 
 
\noindent We have proved one direction of the following 

\smallskip

\index{theorem!Design}
\noindent {\bf Design theorem}. {\sl The set of unit vectors $\{ |\Psi\rangle\}_{I=1}^K$ 
forms a $t$--design if and only if eq. (\ref{designtheorem}) holds.} 

\smallskip

\noindent In the other direction a little more thought is needed \cite{KlaRoe05}. 
Take any vector $|\Psi\rangle$ in ${\mathbbm C}^N$ and construct a vector in 
$({\mathbbm C}^{N})^{\otimes t}$ by taking the tensor product of the 
vector with itself $t$ times. Do the same with $|\bar{\Psi}\rangle$, a vector whose components 
are the complex conjugates of the components, in a fixed basis, of the given vector. A final 
tensor product leads to the vector
\begin{equation} |\Psi\rangle^{\otimes t}\otimes |\bar{\Psi}\rangle^{\otimes t} \in 
({\mathbbm C}^{N})^{\otimes 2t} 
\ . \end{equation}

\noindent In the given basis the components of this vector are 
\begin{equation} (z_0\dots z_0\bar{z}_0\dots \bar{z}_0, z_0\dots z_0\bar{z}_1\dots \bar{z}_1, \dots \dots , 
z_n\dots z_n\bar{z}_n\dots \bar{z}_n) \ . \end{equation}

\noindent In fact the components consists of all possible monomials in Hom$(t,t)$. Thus, to show that 
a set of unit vectors forms a $t$--design it is enough to show that the vector 
\begin{equation} |\Phi\rangle = \frac{1}{K}\sum_I|\Psi_I\rangle^{\otimes t}\otimes 
|\bar{\Psi}_I\rangle^{\otimes t} - \int {\rm d}\Omega_{\Psi}|\Psi\rangle^{\otimes t}\otimes 
|\bar{\Psi}\rangle^{\otimes t} \end{equation}

\noindent is the zero vector. This will be so if its norm vanishes. We observe preliminarily that 
\begin{equation} \langle \Psi_I^{\otimes t}|\Psi_J^{\otimes t}\rangle = \langle \Psi_I|\Psi_J\rangle^t 
\ . \label{designhjalp} \end{equation}

\noindent If we make use of the ubiquitous eq. (7.69)
we find precisely that 
\begin{equation} ||\Phi ||^2 = \frac{1}{K^2}\sum_{I,J}|\langle \Psi_I|\Psi_J\rangle |^{2t} - 
\frac{t!(N-1)!}{(N-1+t)!} \ . \end{equation}

\noindent This vanishes if and only if eq. (\ref{designtheorem}) holds. But 
$|\Phi\rangle = 0$ is a sufficient condition for a $t$--design, and the 
theorem is proven. 

This result is closely related to the {\it Welch bound} \cite{Wel74}, 
which holds for every collection of $K$ vectors in 
${\mathbb C}^N$. For any positive integer $t$ 
\begin{equation} {{N+t-1}\choose{t}} \sum_{I,J}|\langle x_I|x_J\rangle |^{2t} \geq 
\left( \sum_I \langle x_I|x_I\rangle^t\right)^2 \ . \end{equation}

\noindent Evidently a collection of unit vectors forms a $t$-design if and only if 
the Welch bound is saturated. 
The binomial coefficient occurring here is the number of ways in which $t$ identical objects 
can be distributed over $N$ boxes, or equivalently it is the dimension of the symmetric 
subspace ${\cal H}_{\rm sym}^{\otimes t}$ of the $t$--partite Hilbert space 
${\cal H}_N^{\otimes t}$. This is not by accident. Introduce the operator 
\begin{equation} F = \sum_I|\Psi_I^{\otimes t}\rangle \langle \Psi_I^{\otimes t}| \ . \end{equation}

\noindent It is then easy to see, keeping eq. (\ref{designhjalp}) in mind, that 
\begin{equation} \mbox{Tr}F = \sum_I\langle \Psi_I^{\otimes t}|\Psi_I^{\otimes t}\rangle = K 
\end{equation}
\begin{equation} \mbox{Tr}F^2 = \sum_{I,J}\langle \Psi_I^{\otimes t}|
\Psi_J^{\otimes t}\rangle \langle \Psi_J^{\otimes t}|\Psi_I^{\otimes t}\rangle  = 
\sum_{I,J}|\langle \Psi_I|\Psi_J\rangle|^{2t} \ . \end{equation}

\noindent Now we can minimize Tr$F^2$ under the constraint that Tr$F = K$. In fact this 
means that all the eigenvalues $\lambda_i$ of $F$ have to be equal, namely equal to 
\begin{equation} \lambda_i = \frac{K}{\mbox{dim}({\cal H}^{\otimes t}_{\rm sym})} \ . \end{equation}

\noindent So we have rederived the inequality
\begin{equation} \sum_{I,J}|\langle \Psi_I|\Psi_J\rangle|^{2t} = \mbox{Tr}F^2 \geq 
\frac{K^2}{\mbox{dim}({\cal H}^{\otimes t}_{\rm sym})} \ . \end{equation}

\noindent Moreover we see that the operator $F$ projects onto the symmetric subspace.

Although $t$--designs exist in all dimensions, for all $t$, it is not so easy to find 
examples with small number of vectors. A lower bound on the number of vectors needed is 
\cite{Hog82}  
\begin{equation} \mbox{(number of vectors)} \geq \left( \begin{array}{c} N+(t/2)_+ -1 \\ 
(t/2)_+ \end{array} \right) \left( \begin{array}{c} N+(t/2)_- - 1 \\ (t/2)_- 
\end{array} \right) \ , \end{equation}

\noindent where $(t/2)_+$ is the smallest integer not smaller than $t/2$ and 
$(t/2)_-$ is the largest integer not larger than $t/2$. The design is said to be 
{\it tight} if the number of its vectors saturates this bound. Can the bound 
be achieved? For dimension $N = 2$ much is known \cite{HaSl96}. A tight 2--design 
is obtained by inscribing a regular tetrahedron in the Bloch sphere. A tight 3--design is 
obtained by inscribing a regular octahedron, and a tight 5--design by inscribing a regular 
icosahedron. The icosahedron is also the smallest 4--design, so tight 4--designs do not
exist in this dimension. A cube gives a 3-design and a dodecahedron gives a 5-design. 
For dimensions $N > 2$ it is known that tight $t$--designs can exist at most for $t=1,2,3$. 
Every orthonormal basis is a tight 1--design. A tight 2--design needs $N^2$ vectors, and 
the question whether they exist is the subject of Section \ref{sec:SICs}.

 Meanwhile 
we observe that that the $N(N+1)$ vectors in a complete set of MUB saturate 
the Welch bound for $t = 1,2$. Hence complete sets of MUB are 2--designs, and much of 
their usefulness stems from this fact. Tight 3--designs 
exist in dimensions 2, 4, and 6. In general it is not known how many vectors that 
are needed for minimal $t$--designs in arbitrary dimensions, which is why the 
terminology `tight' is likely to be with us for some time.
A 
particularly nice account of all these matters is in the University of 
Waterloo Master's thesis by Belovs (2008) \cite{Bel08}. 
For more results, and references that we have omitted, see Scott \cite{Sco06}. 

The name `design' is used for more than one concept. One example, closely related 
to the one we have been discussing, is that of a 
{\it unitary t--design}.
(Although there was a prehistory, the name seems to stem from 
a University of Waterloo Master's thesis by Dankert (2005) \cite{Dan05}. The idea was further 
developed in papers to which we refer for proofs, applications, and details \cite{GrAuEi07, RS09}).

By definition this is a set of unitary 
operators $\{ U_I\}_{I=1}^K$ with the property that 
\begin{equation} \frac{1}{K}\sum_IU_I^{\otimes t}A(U_I^{\otimes t})^\dagger = \int_{U(N)} {\rm d}U 
\ U^{\otimes t}A(U^{\otimes t})^\dagger \ , \end{equation}

\noindent where $A$ is any operator acting on the $t$--partite Hilbert space and ${\rm d}U$ is 
the normalized Haar measure on the group manifold. In the particularly interesting case $t = 2$ 
the averaging operation performed on the right hand side is known as {\it twirling}.
Condition (\ref{designtheorem}) for when a collection of vectors forms a projective $t$--design 
has a direct analogue: the necessary and sufficient condition for a collection of $K$ unitary 
matrices to form a unitary $t$--design is that  
\begin{equation} \frac{1}{K^2} \sum_{I,J}|\mbox{Tr}U_I^\dagger U_J|^2 = \int_{U(N)} {\rm d}U 
|\mbox{Tr}U|^{2t} = \left\{ \begin{array}{lcl} \frac{(2t)!}{t!(t+1)!} \ , & & N = 2 \\ 
t! \ , & & N \geq t \ . \end{array} \right. \label{Karols} \end{equation}

\noindent When $t > N$ the right hand side looks more complicated. 

It is natural to ask for the 
operators $U_I$ to form a finite group. The criterion for (a projective unitary representation 
of) a finite group to serve as a unitary $t$--design is that it should have the same number of 
irreducible components in the $t$--partite Hilbert space as the group $U^{\otimes t}$ itself. 
Thus a nice error basis, such as the Weyl--Heisenberg group, is always a 1--design because 
any operator commuting with all 
the elements of a nice error basis is proportional to the identity matrix. When $t = 2$ the 
group $U\otimes U$ splits the bipartite Hilbert space into its symmetric and its anti-symmetric 
subspace. 

In prime power dimensions both the Clifford group and the restricted Clifford group 
are unitary 2-designs \cite{DiVLeTe02}. In fact, 
it is enough to use a particular subgroup 
of the Clifford group \cite{Cha05}. 
For qubits, the minimal unitary 2-design is the 
tetrahedral group, which has only 12 elements. In even prime power dimensions $2^k$ 
the Clifford group, but not the restricted Clifford group, is a unitary 3--design 
as well \cite{KuGr15, Web15, Zhu15}. Interestingly, every 
orbit of a group yielding a unitary $t$--design is a projective $t$--design. 
This gives an alternative proof that a complete set of MUB is a 2--design 
(in those cases where it is an  orbit under the restricted Clifford group). 
In even prime power dimensions the set of all 
stabilizer states is a 3--design.
 In dimension 4 it consists of 60 vectors, while a tight 
3--design (which actually exists in this case) has 40 vectors only.

\section{SICs}
\label{sec:SICs}
 
At the end of Section 8.4 
we asked the seemingly innocent question: 
Is it possible to inscribe a regular simplex of full dimension into the 
convex body of density matrices? A tight 2--design in dimension $N$, if it exists, 
has $N^2$ vectors only, and our question can be restated as: Do tight 2--designs 
exist in all dimensions? 
In Hilbert space language the question is: Can we find an informationally complete POVM 
made up of {\it equiangular vectors}? Since absolute values of the scalar products are taken 
the word `vector' really refers to a ray (a point in ${\mathbbm C}{\bf P}^n$). That is, 
we ask for $N^2$ vectors $|\psi_I\rangle$ such that 
\begin{equation} \frac{1}{N}\sum_{I=1}^{N^2}|\psi_I\rangle \langle \psi_I| = {\mathbbm 1} 
\label{sic1} \end{equation} 

\begin{equation} |\langle \psi_I|\psi_J\rangle |^2 = \left\{ \begin{array}{lll} 
1 & \mbox{if} & I = J \\ \\ \frac{1}{N+1} & \mbox{if} & I \neq J \ . \end{array} 
\right. \label{sic2} \end{equation}

\noindent 
We need $N^2$ unit vectors to have informational completeness (in the sense of 
Section 10.1), 
and we are assuming that the mutual fidelities are equal. 
The precise number $1/(N+1)$ follows by squaring 
the expression on the left hand side of eq. (\ref{sic1}), and then taking the trace. Such a collection 
of vectors is called a SIC, so the final form of the question is: Do SICs 
exist?
(The acronym is short for Symmetric Informationally 
Complete Positive Operator Valued Measure \cite{RBSC04}, and is rarely spelled out. We 
prefer to use `SIC' as a noun. When pronounced as `seek' it serves to remind 
us that the existence problem may well be hiding their most important message).

If they exist, SICs have some desirable properties. First of all they saturate 
the Welch bound, and hence they are 2-designs with the minimal number 
of vectors. Moreover, also for other reasons, they are theoretically 
advantageous in quantum state tomography \cite{Sco06}, and they provide a 
preferred form for informationally complete POVMs. Indeed an entire philosophy 
can be built around them \cite{FuSc13}.

But the SIC existence problem is unsolved. 
Perhaps we should begin by noting that there is a crisp non-existence result 
for the real Hilbert spaces ${\mathbbm R}^N$. 
Then Bloch space has dimension $N(N+1)/2-1$, so the number of equiangular 
vectors in a real SIC is $N(N+1)/2$. For $N = 2$ the rays of a real SIC 
pass through the vertices of a regular triangle, and for $N = 3$ the 
six diagonals of an icosahedron will serve. However, for $N > 3$ it 
can be shown that a real SIC cannot exist unless $N+2$ is a square of 
an odd integer. In particular $N = 4$ is ruled out. In fact SICs do not 
exist in real dimension $47 = 7^2-2$ either, so there are further 
obstructions. The non-existence result is due to 
Neumann, and reported by Lemmens and Seidel (1973) \cite{LeSe73}. 
Since then more has been learned \cite{STDH07}. Incidentally the SIC 
in ${\mathbbm R}^3$ has been proposed as an ideal configuration for 
an interferometric gravitational wave detector \cite{Boy10}.

In ${\mathbbm C}^N$ exact solutions are available in all 
dimensions $2\leq N \leq 21$, 
and in a handful of dimensions higher than that. Numerical solutions to high 
precision are available in all dimensions given by two-digit numbers and a bit 
beyond that. (Most of these results, many of them unpublished, are due 
to Gerhard Zauner, Marcus Appleby, Markus Grassl, and Andrew Scott. For the 
state of the art in 2009, see Scott and Grassl \cite{ScGr10}. The first two 
parts of the conjecture are due to Zauner (1999) \cite{Zau99}, the third to 
Appleby at al. (2013) \cite{AYZ13}. We are restating it a little for 
convenience).
 The existing solutions support a three-pronged conjecture:

\

\noindent 1. In every dimension there exists a SIC which is an orbit of the Weyl--Heisenberg group.

\smallskip

\noindent 2. Every vector belonging to such a SIC is invariant under a Clifford group 

element of order 3.

\smallskip 

\noindent 3. When $N > 3$ the overlaps of the SIC vectors are algebraic units in 
an abelian 

extension of the real quadratic field ${\mathbbm Q}(\sqrt{(N-3)(N+1)})$.

\

\noindent Let us sort out what this means, beginning with the easily understood part 1. 

In two dimensions a SIC forms a tetrahedron inscribed in the Bloch sphere. If we orient it so that its 
corners lie right on top of the faces of the octahedron whose corners are the stabilizer states it is easy to see (look at Figure \ref{fig:mub1}a) that the Weyl--Heisenberg 
group can be used to reach any 
corner of the tetrahedron starting from any fixed corner. In other words, when $N = 2$ 
we can always write the $N^2$ SIC vectors in the form 
\begin{equation} |\psi_{r,s}\rangle = D_{r,s}|\psi_0\rangle \ , \hspace{8mm} 0 \leq 
r,s < N-1 \ , \end{equation}

\noindent where $|\psi_0\rangle$ is known as the {\it fiducial vector} for the SIC (and has 
to be chosen carefully, in a fixed representation of the Weyl--Heisenberg 
group). Conjecture 1 says that it is possible to find such a fiducial vector in every 
dimension. Numerical searches are based on this, and basically proceed by minimizing the function 
\begin{equation} f_{\rm SIC} = \sum_{r,s}\left( |\langle \psi_0|\psi_{r,s}\rangle|^2 
- \frac{1}{N+1}\right)^2 \ , \end{equation}

\noindent where the sum runs over all pairs $(r,s) \neq (0,0)$. The arguments of the 
function are the components of the fiducial vector $|\psi_0\rangle$. This is a 
fiducial vector for a SIC if and only if $f_{\rm SIC} = 0$. Solutions have been found 
in all dimensions that have been looked at---even though the presence 
of many local minima of the function makes the task difficult. SICs arising 
in this way are said to be {\it covariant} under the Weyl--Heisenberg group. 
It is believed that the numerical searches for such WH-SICs are exhaustive up to 
dimension $N = 50$ \cite{ScGr10}. They necessarily fall into orbits of the 
Clifford group, extended to include anti-unitary symmetries.
\index{group!extended Clifford} 
For $N \leq 50$ there are six cases where there is only one such orbit 
(namely $N =2$, 4, 5, 10, or 22), 
while as many as ten orbits occur in two cases ($N = 35$ or 39). 

Can SICs not covariant under a group exist? The only publicly available answer to 
this question is that if $N \leq 3$ then all SICs are orbits under the 
Weyl--Heisenberg group \cite{HuSa15}. Can any other group serve the purpose? 
If $N = 8$ there exists an elegant SIC covariant under $H(2)^{\otimes 3}$ 
\cite{Hog98}, as well as two Clifford orbits of SICs covariant under the 
Weyl--Heisenberg group $H(8)$. No other examples of a SIC not covariant 
under $H(N)$ are known, and indeed it is known that for prime $N$ 
the Weyl--Heisenberg group is the only group that can yield SICs \cite{Zhu10}. 
Since the mutually unbiased bases rely on the multipartite Heisenberg group 
this means that there can be no obvious connection between MUB and SICs, 
except in prime dimensions. In prime dimensions it is known that the 
Bloch vector of a SIC projector, when projected onto any one of the MUB 
eigenvalue simplices, has the same length for all the $N+1$ simplices defined 
by a complete set of MUB \cite{ADF14, Kh08}. If $N = 2$, $3$, every state having this 
property is a SIC fiducial \cite{ABBD15}, but when $N \geq 5$ this is far from 
being the case. The two lowest dimensions have a very special status. 

The second part of the conjecture is due to Zauner. It clearly holds if $N = 2$. 
Then the Clifford group, the group that permutes the stabilizer states, is the 
symmetry group of the octahedron. This group contains elements of order 3, and 
by inspection we see that such elements leave some corner of the SIC-tetrahedron 
invariant. The conjecture says that such a symmetry is shared by all SIC vectors 
in all dimensions, and this has been found to hold true for every solution found so far. The sizes of the Clifford orbits shrink accordingly. In many dimensions---very 
much so in 19 and 48 dimensions---there are SICs 
left invariant by larger subgroups of the Clifford group, but the order 3 
symmetry is always present and appears to be universal. 
There is no understanding of why this should be so. 

To understand the third part of the conjecture, and why it is interesting, 
it is necessary to go into the methods used to produce solutions in the first place.  
Given conjecture 1, the straightforward way to find a SIC is to solve the 
equations  
\begin{equation} |\langle \psi_0|D_{rs}|\psi_0\rangle|^2 =  
\frac{1}{N+1} 
\end{equation}

\noindent for $(r,s) \neq (0,0)$. Together with the normalization this is a set 
of $N^2$ multivariate quartic polynomial equations in the $2N$ 
real variables needed to specify the fiducial vector. To solve them 
one uses the method of Gr\"obner bases to reduce the set of equations to a single 
polynomial equation in one variable \cite{ScGr10}. This is a task for computer 
programs such as MAGMA, and a clever programmer. The number of equations greatly 
exceeds the number of variables, so it would not be surprising if they did not have a 
solution. But solutions do exist. 

As an example, here is a fiducial vector for a SIC in 4 
dimensions \cite{RBSC04, Zau99}:
\begin{equation} \psi_0 =  \left( \begin{array}{c} 
\frac{1-\tau}{2}\sqrt{\frac{5+\sqrt{5}}{10}} \\ 
\frac{1}{20}\left( i\sqrt{50-10\sqrt{5}} + (1+i)\sqrt{5(5+3\sqrt{5}}\right) \\
-\frac{1+\tau}{2}\sqrt{\frac{5+\sqrt{5}}{10}} \\
\frac{1}{20}\left( i\sqrt{50-10\sqrt{5}} - (1-i)\sqrt{5(5+3\sqrt{5}}\right)
\end{array} \right)  \label{SICfid}
\end{equation}

\noindent (with $\tau = -e^{i\pi/4}$). This does not look memorable at first sight. 
Note though that all components can be expressed in terms of nested square roots. 
This means that they are numbers that can be constructed by means of rulers and 
compasses, just as the ancient Greeks would have wished. This is not at all what 
one would expect to come out from a Gr\"obner basis calculation, which in the end 
requires one to solve a polynomial equation in one variable but of high degree. 
Galois showed long ago that generic polynomial equations cannot be solved 
by means of nested root extractions if their degree exceeds four. 

And we can simplify the expression. Using the fact that the Weyl--Heisenberg 
group is a unitary operator basis, eqs. (\ref{WHUOB1}-\ref{WHUOB2}) we can 
write the fiducial projector as 
\begin{equation} |\psi_0\rangle \langle \psi_0| = \frac{1}{N}\sum_{r,s=0}^{N-1} 
D_{r,s} \langle \psi_0|D_{r,s}^\dagger |\psi_0\rangle \ . \end{equation}

\noindent But the modulus of the overlaps is fixed by the SIC conditions, so 
we can define the phase factors 
\begin{equation} e^{i\theta_{r,s}} = \sqrt{N+1} \langle 
\psi_0|D_{r,s}|\psi_0\rangle \ , \hspace{8mm} (r,s) \neq (0,0) \ .  \end{equation}

\noindent These phase factors are independent of the choice of basis in ${\mathbb C}^N$, 
and if we know them we can reconstruct the SIC. The 
number of independent phase factors is limited by the Zauner symmetry (and 
by any further symmetry that the SIC may have). For the $N = 4$ example we find 
\begin{equation} \left[ \begin{array}{cccc} \times & e^{i\theta_{0,1}} & e^{i\theta_{0,2}} 
& e^{i\theta_{0,3}} \\ e^{i\theta_{1,0}} & e^{i\theta_{1,1}} & e^{i\theta_{1,2}} 
& e^{i\theta_{1,3}} \\ e^{i\theta_{2,1}} & e^{i\theta_{2,1}} & e^{i\theta_{2,2}} 
& e^{i\theta_{2,3}} \\ e^{i\theta_{3,0}} & e^{i\theta_{3,1}} & e^{i\theta_{3,2}} 
& e^{i\theta_{3,3}} \end{array} \right]  = \left[ 
\begin{array}{rrrr} \times & u &-1 & 1/u \\ 
u & 1/u & -1/u & 1/u \\ -1 & -u & -1 & 1/u \\ 1/u & u & u & u \end{array} 
\right] , \label{sicfas} \end{equation}

\noindent where 
\begin{equation} u = \frac{\sqrt{5}-1}{2\sqrt{2}}+\frac{i\sqrt{\sqrt{5}+1}}{2}
\ . \label{u} \end{equation}

\noindent The pattern in eq. (\ref{sicfas}) is forced upon us by the Zauner symmetry, 
so once we know the number $u$ it is straightforward to reconstruct the entire SIC. 

What is this number? To answer this question one computes (usually by means of 
a computer) the {\it minimal polynomial}, the lowest degree polynomial with 
coefficients among the integers satisfied by the number $u$. In this case it is
\begin{equation} p(t) = t^8 -2t^6 - 2t^4 - 2t^2 +1 \ . \label{minpol} \end{equation}

\noindent Because the minimal polynomial exists, we say that $u$ is an 
{\it algebraic number}. Because its leading coefficient equals 1, we say that $u$ 
is an {\it algebraic integer}. Algebraic integers form a ring, just as the ordinary 
integers do. (Algebraic number theorist refer to ordinary integers as 
`rational integers'. This is the special case when the polynomial is of first 
order.) Finally we observe that $1/u$ is an algebraic integer too---in fact it 
is another root of the same equation---so we say that $u$ is an {\it algebraic 
unit}.
Neither algebraic number theory nor Galois theory (which we are 
coming to) lend themselves to thumbnail sketches. For first encounters we recommend 
the books by Alaca and Williams \cite{AW04} and by Howie \cite{Ho06}, respectively.

Hence $u$ is a very special number, and the question arises what number field it 
belongs to. This is obtained by adding the roots of (\ref{minpol}) to the field of 
rational numbers, ${\mathbb Q}$. We did give a thumbnail sketch of field extensions 
in Section \ref{sec:primepower}, but now we are interested in fields with an 
infinite number of elements. When we adjoined a root of the equation $x^2 +1 = 0$ 
to the real field ${\mathbbm R}$ we obtained the complex field ${\mathbbm R}({\rm i}) = 
{\mathbbm C}$. In fact there are two roots of the equation, and there is a group---in 
this case the abelian group $Z_2$---that permutes them. This group is called the 
{\it Galois group} 
of the extension. It can also be regarded as the group of automorphisms of the extended field 
${\mathbbm C}$ that leaves the ground field ${\mathbbm R}$ invariant. A Galois group arises 
whenever a root of an irreducible polynomial is added to a field. Galois proved that 
a polynomial with rational coefficients can be solved with root extractions if and only 
if the Galois group is soluble 
\index{group!soluble}
(as is the case for a generic quartic, but not for a generic quintic). This is 
the origin of the name `soluble'. For a group to be soluble it must have a particular 
pattern of invariant subgroups. 

What one finds in the case at hand is that the Galois group is non-abelian, but only barely 
so. The field for the $N = 4$ SIC can be obtained by first extending ${\mathbb Q}$ to 
${\mathbb Q}(\sqrt{5})$, that is to say to consider the {\it real quadratic field} 
consisting of all numbers of the form $x + \sqrt{5}y$, where $x,y$ are rational. A further 
extension then leads to the field ${\mathbb Q}(u,r)$, where $r$ is an additional (real) 
root of (\ref{minpol}). The Galois group arising in the second step, considered by itself, 
is abelian, and the second extension is therefore said to be an {\it abelian extension}.
  
Thus the mysterious number $u$ does not only belong to the field whose construction we sketched, 
it has a very special status in it, and moreover the field is of an especially important 
kind, technically known as a ray class field.

Things are not quite so simple in higher dimensions, but almost so---in 
terms of principles that is, not in terms of the calculations that need to be 
done.
(The number theory of SICs, so far as it is known, was developed by 
Appleby, Flammia, McConnell, and Yard (2016) \cite{AFMY16}). 
The SIC overlaps, and the SIC vectors themselves if expressed in the natural basis, are still 
given by nested radicals, although no longer by square roots only so they are not 
constructible with ruler and compass. Indeed the third part of the SIC conjecture 
holds in all cases that have been looked at. What is more, the SIC overlaps 
continue to yield algebraic units. But it is not understood why the polynomial 
equations that define the SICs have this property. 

A field extension is said to be abelian whenever its Galois group is abelian. 
In the nineteenth century Kronecker and Weber studied 
abelian extensions of the rational field ${\mathbbm Q}$, and they proved that all such 
extensions are subfields of a {\it cyclotomic field}, which is what one obtains by 
adjoining a root of unity to the rational numbers. (It will be observed that MUB vectors 
have all their overlaps in a cyclotomic field.) 
\index{field!cyclotomic}
Kronecker's {\sl Jugendtraum} was to 
extend this result to much more general ground fields. For instance, there are quadratic 
extensions of the rationals such as ${\mathbbm Q}(\sqrt{5})$, consisting of all numbers 
of the form $x_1 + x_2\sqrt{5}$, where $x_1,x_2$ are rational numbers. More generally 
one can replace the square root of two with the square root of any integer. If that integer 
is negative the extension is an imaginary quadratic extension, otherwise it is a real 
quadratic extension. 
For the case of abelian extensions of an imaginary quadratic extension 
of ${\mathbb Q}$ Kronecker's dream led to brilliant successes, with deep connections to---among 
other things---the theory of elliptic curves. Such numbers turn up for special choices of the 
arguments of elliptic and modular functions, much as the numbers in a cyclotomic field turn 
up when the function $e^{i\pi x}$ is evaluated at rational values of $x$.  
As the 12th on his famous list of problems for 
the twentieth century, Hilbert asked for the restriction to imaginary quadratic fields 
to be removed.
(For an engaging account of this piece of history see 
the book by Gray \cite{Gray00}). The natural first step in solving the 12th 
problem would seem to be to find a framework for the abelian extensions 
of the real quadratic fields. This remains unsolved, but it seems to be in these deep waters that the SICs are swimming. 

\bigskip 
           \centerline{ *  *  *  }   
\bigskip 

This chapter may have left the reader a little bewildered, and appalled by equations 
like (\ref{SICfid}). In defence of the chapter, we observe that practical applications 
(to signal processing, adaptive radar, and more) lie close to it. But the last word goes 
to Hilbert \cite{Re70}:  

\smallskip
\noindent {\sl There still lies an abundance of priceless treasures hidden in this domain, belonging 
as a rich reward to the explorer who knows the value of such treasures and, with love, 
pursues the art to win them.} 
\smallskip

\section{Concluding remarks}

\vspace{5mm}

The aim of these notes is literally to present
a concise introduction to the broad subject of
discrete structures in a finite Hilbert space.
We are convinced that such a knowledge is useful 
while investigating numerous problems motivated by 
the theory of quantum information processing.
Even for small dimensions of the Hilbert space
several intriguing questions remain open,
so we are pleased to encourage the 
reader to contribute to this challenging field.

\bigskip

We are indebted to  Marcus Appleby, 
Dardo Goyeneche, Marcus Grassl,
David Gross and Huangjun Zhu, for reading some fragments 
of the text and providing us with valuable remarks. 

\smallskip

Financial support by Narodowe Centrum Nauki 
under the grant number DEC-2015/18/A/ST2/00274
is gratefully acknowledged.

\appendix

\medskip
\section{Contents of the II edition of the book "Geometry of Quantum States. 
         An Introduction to Quantum Entanglement"
         by {\sl I. Bengtsson and K. {\.Z}yczkowski}}

{\small

 {\bf 1 Convexity, colours and statistics }
 
\quad {1.1} Convex sets                                

\quad {1.2} High dimensional geometry

\quad  {1.3} Colour theory

\quad  {1.4} What is ``distance''? 

\quad  {1.5} Probability and statistics

{\bf 2  Geometry of probability distributions } 

\quad {2.1} Majorization and partial order

\quad {2.2} Shannon entropy

\quad {2.3} Relative entropy

 \quad {2.4} Continuous distributions and measures

\quad {2.5} Statistical geometry and the Fisher--Rao metric

 \quad {2.6} Classical ensembles

\quad {2.7} Generalized entropies

 {\bf 3 Much ado about spheres}
 
\quad {3.1} Spheres

\quad  {3.2} Parallel transport and statistical geometry

 \quad {3.3} Complex, Hermitian, and K\"ahler manifolds

\quad  {3.4} Symplectic manifolds

\quad {3.5} The Hopf fibration of the $3$-sphere

\quad  {3.6} Fibre bundles and their connections

\quad  {3.7} The $3$-sphere as a group

\quad  {3.8} Cosets and all that

 {\bf 4  Complex projective spaces}

\quad  {4.1} From art to mathematics

\quad  {4.2} Complex projective geometry

\quad  {4.3} Complex curves, quadrics and the Segre embedding

\quad {4.4} Stars, spinors, and complex curves

\quad {4.5} The Fubini-Study metric

\quad {4.6} ${\mathbb C}{\bf P}^n$ illustrated

\quad {4.7} Symplectic geometry and the Fubini--Study measure

\quad {4.8} Fibre bundle aspects

\quad {4.9} Grassmannians and flag manifolds

 {\bf 5 Outline of quantum mechanics}

 \quad{5.1} Quantum mechanics

 \quad{5.2} Qubits and Bloch spheres

 \quad{5.3} The statistical and the Fubini-Study distances

 \quad{5.4} A real look at quantum dynamics

 \quad{5.5} Time reversals

 \quad{5.6} Classical \& quantum states: a unified approach

 \quad{5.7} Gleason and Kochen-Specker 

 {\bf 6 Coherent states and group actions }

 \quad{6.1} Canonical coherent states

 \quad{6.2} Quasi-probability distributions on the plane

 \quad{6.3} Bloch coherent states

 \quad{6.4} From complex curves to $SU(K)$ coherent states

 \quad{6.5} $SU(3)$ coherent states

 {\bf 7 The stellar representation}

 \quad{7.1} The stellar representation in quantum mechanics

 \quad{7.2} Orbits and coherent states

 \quad{7.3} The Husimi function

 \quad{7.4} Wehrl entropy and the Lieb conjecture

 \quad{7.5} Generalised Wehrl entropies

 \quad{7.6} Random pure states

 \quad{7.7} From the transport problem to the Monge distance

 {\bf 8 The space of density matrices}

 \quad{8.1} Hilbert--Schmidt space and positive operators

 \quad{8.2} The set of mixed states

 \quad{8.3} Unitary transformations

 \quad{8.4} The space of density matrices as a convex set

\quad{8.5} Stratification

\quad{8.6} Projections and cross--sections 

 \quad{8.7} An algebraic afterthought

 \quad{8.8} Summary

 {\bf 9 Purification of mixed quantum states}

 \quad{9.1} Tensor products and state reduction

\quad{9.2} The Schmidt decomposition

\quad {9.3} State purification \& the Hilbert-Schmidt bundle

\quad {9.4} A first look at the Bures metric

 \quad{9.5} Bures geometry for $N=2$

 \quad{9.6} Further properties of the Bures metric

 {\bf 10 Quantum operations} 

 \quad{10.1} Measurements and POVMs

 \quad{10.2} Algebraic detour: matrix reshaping and reshuffling

 \quad{10.3} Positive and completely positive maps

\quad {10.4} Environmental representations

\quad {10.5} Some spectral properties

\quad{10.6} Unital \& bistochastic maps

\quad {10.7} One qubit maps

 {\bf 11 Duality: maps versus states}

 \quad{11.1} Positive \& decomposable maps

\quad {11.2} Dual cones and super-positive maps

 \quad{11.3} Jamio{\l }kowski isomorphism

 \quad{11.4} Quantum maps and quantum states

 {\bf 12  Discrete structures in Hilbert space} 

\quad{12.1}  Unitary operator bases and the Heisenberg groups

\quad{12.2} Prime, composite, and prime power dimensions

\quad{12.3}  More unitary operator bases 

\quad{12.4} Mutually unbiased bases

\quad{12.5} Finite geometries and discrete Wigner functions

\quad{12.6} Clifford groups and stabilizer states

\quad{12.7} Some designs

\quad{12.8} SICs

{\bf 13  Density matrices and entropies} 

 \quad{13.1} Ordering operators

 \quad{13.2} Von Neumann entropy

\quad {13.3} Quantum relative entropy

\quad {13.4} Other entropies

 \quad{13.5} Majorization of density matrices

\quad{13.6} Proof of the Lieb conjecture

\quad{13.7} Entropy dynamics

 {\bf 14 Distinguishability measures}

\quad{14.1} Classical distinguishability measures

\quad {14.2} Quantum distinguishability measures

\quad {14.3} Fidelity and statistical distance

 {\bf 15 Monotone metrics and measures}

\quad {15.1} Monotone metrics

\quad {15.2} Product measures and flag manifolds

 \quad{15.3} Hilbert-Schmidt measure

 \quad{15.4} Bures measure

 \quad{15.5} Induced measures

 \quad{15.6} Random density matrices

\quad {15.7} Random operations

\quad{15.8} Concentration of measure

 {\bf 16 Quantum entanglement}

 \quad{16.1} Introducing entanglement

 \quad{16.2} Two qubit pure states: entanglement illustrated

 \quad{16.3} Maximally entangled states

 \quad{16.4} Pure states of a bipartite system

\quad{16.5} A first look at entangled mixed states 

\quad{16.6} Separability criteria

\quad{16.7} Geometry of the set of separable states

 \quad{16.8} Entanglement measures

 \quad{16.9} Two qubit mixed states 

{\bf 17 Multipartite entanglement}

 \quad{17.1} How much is three larger than two? 

 \quad{17.2} Botany of states

\quad{17.3} Permutation symmetric states

\quad{17.4} Invariant theory and quantum states 

\quad{17.5} Monogamy relations and global multipartite entanglement

\quad{17.6} Local spectra and the momentum map

\quad{17.7} AME states and error--correcting codes 

\quad{17.8} Entanglement in quantum spin systems

{\bf Epilogue }

\quad  Appendix 1 Basic notions of differential geometry 

\quad Appendix 2 Basic notions of group theory 

\quad Appendix 3  Geometry do it yourself 

\quad Appendix 4  Hints and answers to the exercises
} 

\bigskip


\end{document}